\documentclass{aa}
\usepackage{graphicx}

\begin{document}
\newcommand{\Hy}{\ion{H}}
\newcommand{\Fe}{\ion{Fe}}
\newcommand{\Po}{\ion{P}}
\newcommand{\Ti}{\ion{Ti}}
\newcommand{\fe}{\ion{Fe}{ii}}
\newcommand{\h}{H$_2$}
\newcommand{\ci}{\ion{C}{i}}
\newcommand{\oi}{\ion{O}{i}}
\newcommand{\sii}{\ion{S}{ii}}
\newcommand{\n}{\ion{N}{i}}
\newcommand{\he}{\ion{He}{i}}
\newcommand{\tiii}{\ion{Ti}{ii}}
\newcommand{\coii}{\ion{Co}{ii}}
\newcommand{\pii}{\ion{P}{ii}}
\newcommand{\hi}{\ion{H}}
\title{Near-infrared, IFU spectroscopy unravels the bow-shock HH99B\thanks{Based on observations collected at the European Southern Observatory (La Silla and 
Paranal), Chile (77.C-0203)}}
\author{T.Giannini$^1$, L.Calzoletti$^{1,2}$, B.Nisini$^1$, C.J.Davis$^3$, J.Eisl\"{o}ffel$^4$,
M.D.Smith$^5$}
\offprints{Teresa Giannini, email:giannini@oa-roma.inaf.it}
\institute{{$^1$ INAF-Osservatorio Astronomico di Roma, via Frascati 33, I-00040 Monte Porzio Catone, Italy\\
$^2$ Universit\`a di Cagliari, Via Universit\`a 40, I-09124 Cagliari, Italy\\
$^3$ Joint Astronomy Centre, 660 North A'oh$\overline{o}$k$\overline{u}$ Place, University Park, Hilo, HI 96720, USA\\
$^4$ Th\"{u}ringer Landessternwarte, Sternwarte 5, D-07778 Tautenburg, Germany\\
$^5$ Centre for Astrophysics \& Planetary Science, School of Physical Sciences, 
The University of Kent, Canterbury CT2 7NR, United Kingdom}\\
\email{giannini,nisini,calzol@oa-roma.inaf.it, c.davis@jach.hawaii.edu, jochen@tls-tautenburg.de,
m.d.smith@kent.ac.uk}}
%
%
\date{Received date; Accepted date}
%
%
%
\titlerunning{Bidimensional analysis of H99B}
\authorrunning{T.Giannini et al.}

\abstract{{\it Aims.} We aim to characterise the morphology and the physical parameters governing
the shock physics of the Herbig-Haro object HH99B. We have obtained  
SINFONI-SPIFFI IFU spectroscopy ({\it R} $\sim$ 2000 $-$ 4000) 
between  1.10 and 2.45 $\mu$m detecting more than 170 emission lines, to a large extent never observed before in a Herbig-Haro object.
Most of them come from ro-vibrational transitions of molecular hydrogen (v$_{up}$\,$\le$\,7, E$_{up}\la$
38~000 K) and [\fe] (E$_{up}\la$ 30~000 K). In addition, we observed several hydrogen and helium 
recombination lines, along with fine structure lines of ionic species. All the brightest
lines appear resolved in velocity.\\
{\it Methods.} Intensity ratios of ionic lines have been compared with predictions of NLTE models
to derive bi-dimensional maps of extinction and electron density, along with estimates of
temperature, fractional ionisation and atomic hydrogen post-shock density. H$_2$ line intensities have been
interpreted in the framework of Boltzmann diagrams, from which we have derived maps of 
extinction and temperature of the molecular gas. From the intensity maps of bright
lines (i.e. H$_2$ 2.122$\mu$m and [\fe] 1.644$\mu$m) the kinematical properties of the shock(s)
at work in the region have been delineated.
Finally, from selected [\fe] lines, constraints on the spontaneous emission coefficients of the 
1.257, 1.321 and 1.644 $\mu$m lines are provided.\\
{\it Results.} Visual extinction variations up to 4 mag emerge, showing that the usual assumption
of constant extinction could be critical. The highest A$_V$ is found at the bow-head (A$_V$ $\sim$ 4~mag)
while diminishing along the flanks. 
The electron density increases from $\sim$ 3~10$^3$ cm$^{-3}$ in the receding parts of the shock to 
$\sim$ 6~10$^3$ cm$^{-3}$ in the apex, where we estimate, from [\fe] line ratios, a temperature
of $\sim$ 16~000 K. Molecular gas temperature is lower in the bow-flanks (T$\sim$ 3000 K), then 
progressively increasing toward the head up to T$\sim$ 6000 K.
In the same zone, we are able to derive from the 
[\fe]1.257/[\pii]1.187 line ratio, the iron gas-phase abundance ($\sim$ 60\% of the solar 
value) along with the hydrogen fractional ionisation (up to 50\% at the bow-head) and the 
atomic hydrogen post-shock gas density ($\sim$ 1 10$^4$ cm$^{-3}$).
The kinematical properties derived for the molecular gas substantially confirm those published in 
Davis et al.(1999), while new information (e.g. v$_{shock}$ $\sim$ 115 km s$^{-1}$) is provided 
for the shock component responsible for the ionic emission. We also provide an indirect measure
of the H$_2$ breakdown speed (between 70 and 90 km s$^{-1}$) and compute the inclination angle with respect to the line of sight.
The map parameters, along with images of the observed line intensities, will be used to put 
stringent constraints on up-to-date shock models.
} 
\maketitle  
%
%


\section{Introduction}
Mass loss phenomena in the form of powerful bipolar jets and molecular outflows
are often associated with the early evolution of protostars. Together
with indirectly regulating the accretion process, they play a crucial r\^{o}le also 
in the interaction between the protostar and the natal environment, 
causing injection of  momentum, kinetic energy and turbulence in the ISM,
along with irreversible modifications of its chemical structure and 
physical conditions.
The most violent interaction occurs at the terminal working surface, where 
the supersonic flow impacts the undisturbed medium and 
most of the ambient material is entrained by the jet (e.g. Reipurth \& Bally, 2001).
In a schematic model, the interaction occurs via two shocks: an internal working surface 
(Mach disk) which decelerates the jet gas and a forward shock which accelerates 
the ambient gas producing a mixture of shock velocities (e.g. Hartigan, 1989). 
This latter has often a curve-shaped morphology: therefore, only the component of motion along the
jet axis is slowed going from the apex of the shock toward the receding parts, where
large transverse motions occur. In a classical scenario, for sufficiently fast shocks,
the head of the bow is a pure dissociative (J)ump-type shock (Hollenbach \& McKee, 1989), which changes along the bow flanks, 
where the impact occurs obliquely, to slower, non-dissociative (C)ontinuos-type shocks 
(Draine, 1980). 
In this framework, highly excitated ionic emission should arise at the head of the bow, while
molecular emission mainly originates from the cooling regions along the flanks and behind
the bow. Recent models, however, predict that mixtures of J- and C-type shocks can occur along 
the overall structure of the bow, or, in the presence of a sufficiently strong magnetic
field, that a J-type shock can evolve into a C-type, remaining embedded at early time 
(Smith \& Rosen, 2003; Smith \& Mac Low, 1997; Flower et al., 2003; Le Bourlot et al., 2002, 
Lesaffre et al., 2004a,b).

The large number of parameters which regulate the shock physics (e.g. strength and direction of the local 
magnetic field, ionisation fraction and density gradient between the shocked gas and the local environment)
can be constrained only through dedicated observations aimed at probing the physical and kinematical 
properties of the gas all along the bow structure.
In this respect, observations in the near-infrared represent a well suited tool:
both H$_2$ ro-vibrational lines, which are the main gas coolants in continuos shock 
components, and fine structure transitions of abundant atomic species (e.g. {\ion{Fe}{ii}},
{\ion{C}{i}}, {\ion{S}{ii}}), which are expected to emit in dissociative shocks, fall in this
wavelength range.
Indeed, in recent years multiple observations of jets and bow shocks have been conducted in the
near-infrared by means of long-slit spectroscopy (e.g. Eisl\"{o}ffel, Smith \& Davis, 2000;
Giannini et al., 2004; Smith, Froebrich \& Eisl\"{o}ffel, 2003; Nisini et al., 2002; 
O'Connell, Smith \& Davis, 2004; Nisini et al., 2005). The observed spectra are typically rich in both 
molecular (H$_2$) and ionic ([FeII]) lines, implying that strong gradients in the local conditions 
(e.g. temperature, fractional ionisation) do occur and that multiple shock components are
simultaneously at work.\\ 
The main limitation of long-slit spectroscopy for studying in detail 
bow-shock morphologies arises from the poor coverage of the extended bow surface usually obtainable 
in a reasonable amount of observing time; on the contrary, Integral Field Spectroscopy (IFU) represents a well-tailored tool to overcome this problem, since it 
allows us to obtain simultaneously 2-D maps at different wavelengths. In this paper we present the spectral
images obtained with the IFU facility SINFONI (Eisenhauer et al., 2003, Bonnet et al., 2004)
of a prototype bow-shock, namely the Herbig-Haro object HH99B.
This is located in the RCrA molecular core at d$\sim$ 130 pc (Marraco \& Rydgren, 1981), and it was firstly 
discovered in the optical by Hartigan \& Graham (1987), who suggested it is the red-shifted lobe of 
the outflow powered by the HH100-IR source. More recently, Wilking et al.(1997) have proposed, on the basis of 
near-infrared images, the infrared source IRS9 and the Herbig Ae star RCrA as other possible
exciting source candidates.  
HH99B was firstly imaged in the near-infrared by Davis et al. (1999, hereafter D99) who
identified three different emission zones: one at the head of the bow (subsequently named B0 by
M$^c$Coey et al., 2004, hereafter MC04) where the bulk of the 
emission comes from ionised gas, and two bow-flanks (knots B1 and B3), which emit mainly in H$_2$ lines.
A further H$_2$ knot, immediately behind the bow apex, was identified as knot B2. 
In the framework of both bow- (D99) and planar- (MC04) shock models, 
some attempt has been made to model the line emission observed in HH99B: the H$_2$ morphology is well 
fitted by a C-type bow shock, while the physical conditions of the molecular
gas (measured  by means of long-slit spectroscopy) have been accounted for by a planar
J-type shock with a magnetic precursor. None of these two models, however, is able to reproduce
the copious ionic emission, that requires the presence of a further fully dissociative shock component.\\
With the present work we aim at putting strong observational constraints on bow-shock models.
(whose detailed application will be subject of a separated paper).  
In particular we intend to: {\it (i)} morphologically characterise the emission of the different lines; {\it (i)} derive maps of 
the main physical parameters which govern the shock physics; {\it (iii)} study the velocity field(s) 
along the bow structure.\\

Our work is organised as follows: Section 2 describes our observations and 
the obtained results; the line excitation analysis and the kinematical properties are then presented 
in Sections 3 and 4. Concluding remarks are given in Section 5, while in Appendix A we describe the
procedure we have applied to derive the inclination angle of HH99B.

\section{Observations and Results}

HH99B was observed during four different runs in May and July 2006 with the SINFONI-SPIFFI instrument
at the VLT-UT4 (ESO Paranal, Chile). The coordinates of the pointed position are: $\alpha_{2000}$=19$^h$02$^m$05.4$^s$,
$\delta_{2000}$=$-$36$^{\circ}$54$^{\prime}$39$^{\prime\prime}$. 
The Integral Field Unit was employed to obtain spectroscopic data in J (1.10-1.40 $\mu$m), H (1.45-1.85 $\mu$m) and K 
(1.95-2.45 $\mu$m) bands, at spectral resolution 2000, 3000, and 4000, respectively, each
wavelength band fitting on the 2048 pixels of the Hawaii detector in the dispersion
direction. An image slicer converts the bidimensional field-of-view into a one-dimensional long-slit, 
that is fed into a spectrograph to disperse the light of each pixel simultaneously. As a consequence 
seeing and atmospheric response can affect neither
the emission morphology at a given wavelength nor intensity ratios of lines within the same filter.
  The spatial resolution was 
selected at 0.25$^{\prime\prime}$ per image slice, which corresponds to a field-of-view 
of 8$^{\prime\prime} \times$8$^{\prime\prime}$. No adaptive optics is supported in this configuration.
The total integration time is 2400~s, 1500~s, and 1800~s in J, H, and K bands, respectively.\\
The observations were acquired by nodding the telescope in the usual ABB$^{\prime}$A$^{\prime}$ mode 
and a telluric B-type standard star close to the source was observed to remove the atmospheric spectral
response.\\

\begin{figure*}
\centering
\includegraphics[width=14cm]{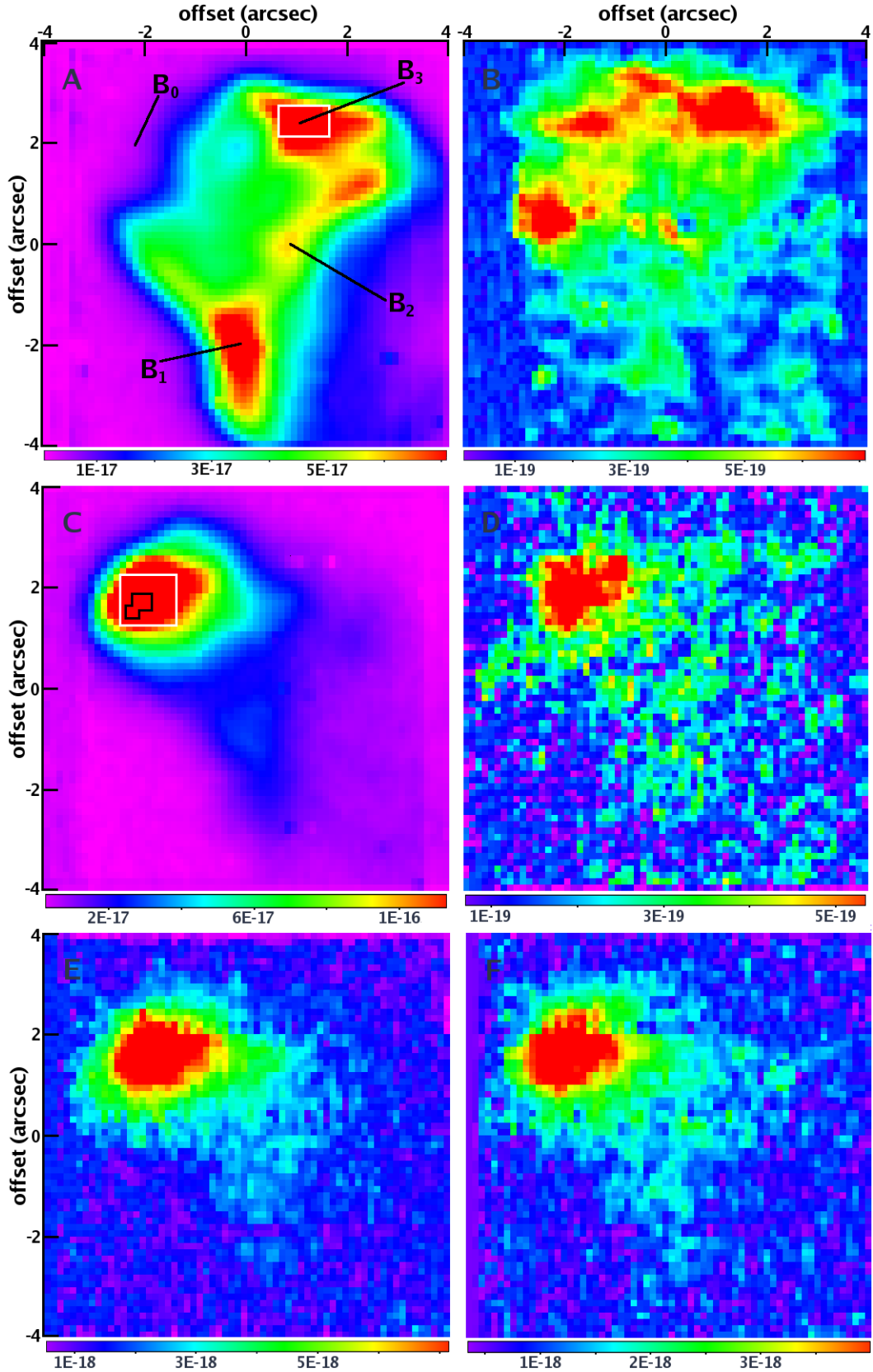}
\caption{\label{fig:righe} Selected spectral images from our HH99B data-cube. Intensities are given in colour 
scale. Offsets are from $\alpha_{2000}$=19$^h$02$^m$05.4$^s$, 
$\delta_{2000}$=$-$36$^{\circ}$54$^{\prime}$39$^{\prime\prime}$. A) H$_2$:1-0 S(1) 
at 2.122$\mu$m. The locations of the knots labelled by D99 (B1, B2 and B3) and MC04 (B0) 
are indicated; B) H$_2$: 2-1S(17) at 1.758$\mu$m; C) [\fe]: a$^4D_{7/2}$-a$^4F_{9/2}$ at 1.644$\mu$m. 
The black line delimits the area where [\fe] lines at S/N $>$ 100  are detected and used to construct 
the plot of Figure\,\ref{fig:av}; D) [\fe]: a$^4P_{3/2}$-a$^4D_{7/2}$ at 1.749$\mu$m; E) H:Pa$\beta$ at 
1.282$\mu$m; F) [\pii]: $^2D_{2}$-$^3P_{2}$ at 1.188$\mu$m. White rectangles in panels A) and C) indicate the areas over which we have integrated 
the signal and extracted the spectra shown in Figures\,\ref{fig:spettro_j}-\ref{fig:spettro_k}.}
\end{figure*}

\begin{figure*}
\centering
\includegraphics[width=15cm]{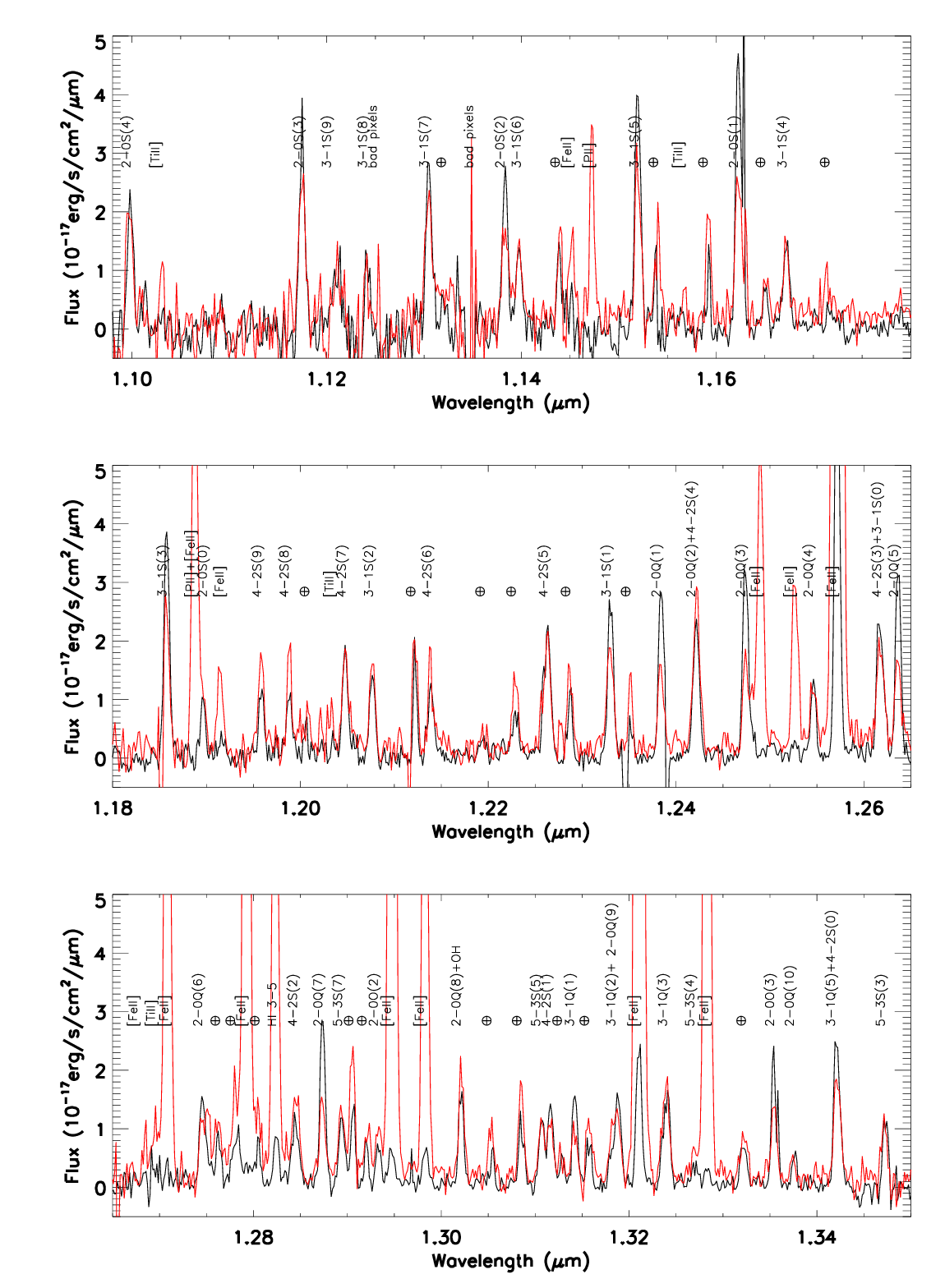}
\caption{\label{fig:spettro_j} J band spectrum extracted in knot B3 (black) and
B0 (red), over an area of 0.5 and 1.25 arcsec$^2$, respectively.
}
\end{figure*}

\begin{figure*}
\centering
\includegraphics[width=15cm]{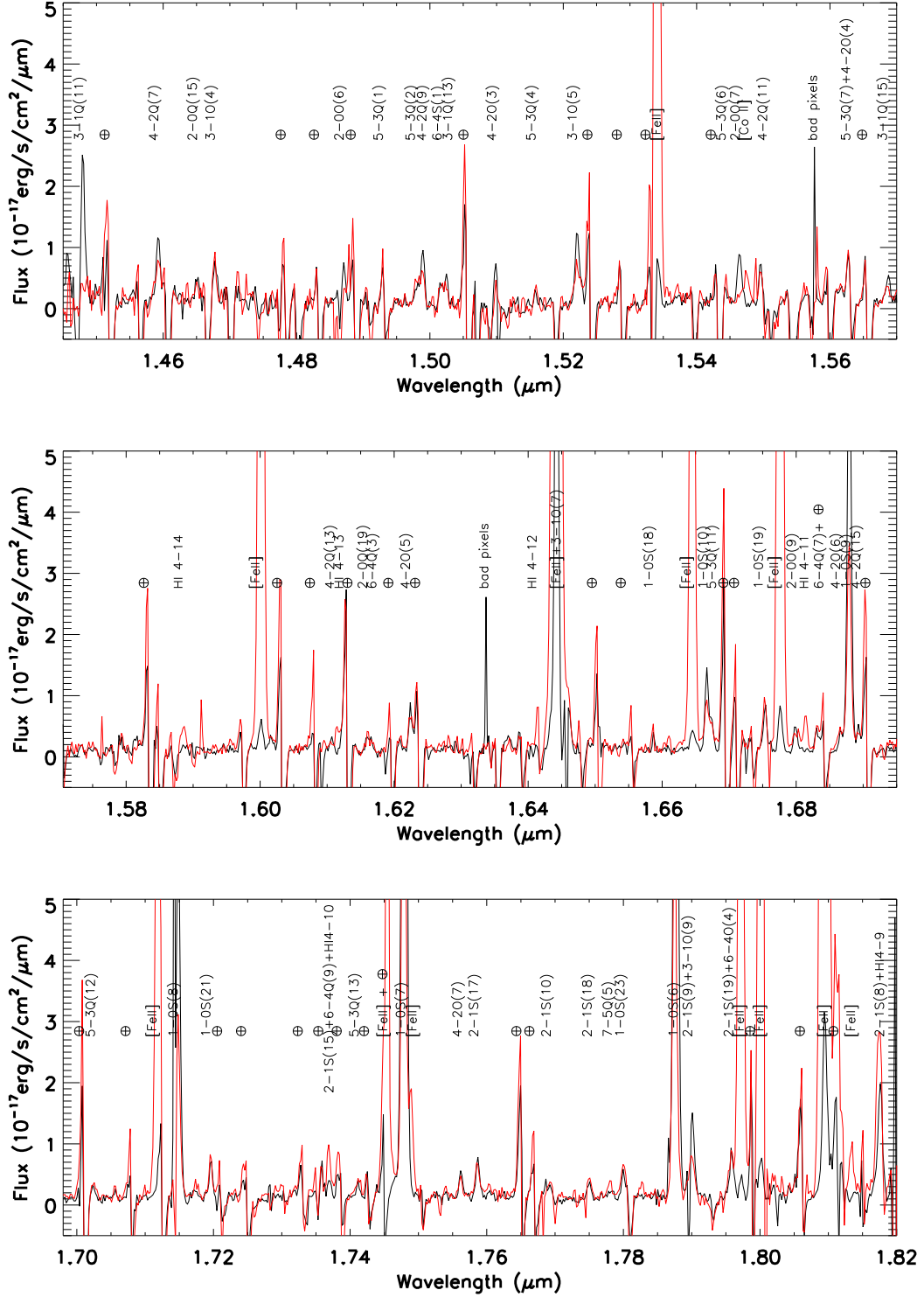}
\caption{\label{fig:spettro_h} As in Figure\, \ref{fig:spettro_j} for the H band.}
\end{figure*}

\begin{figure*}
\centering
\includegraphics[width=15cm]{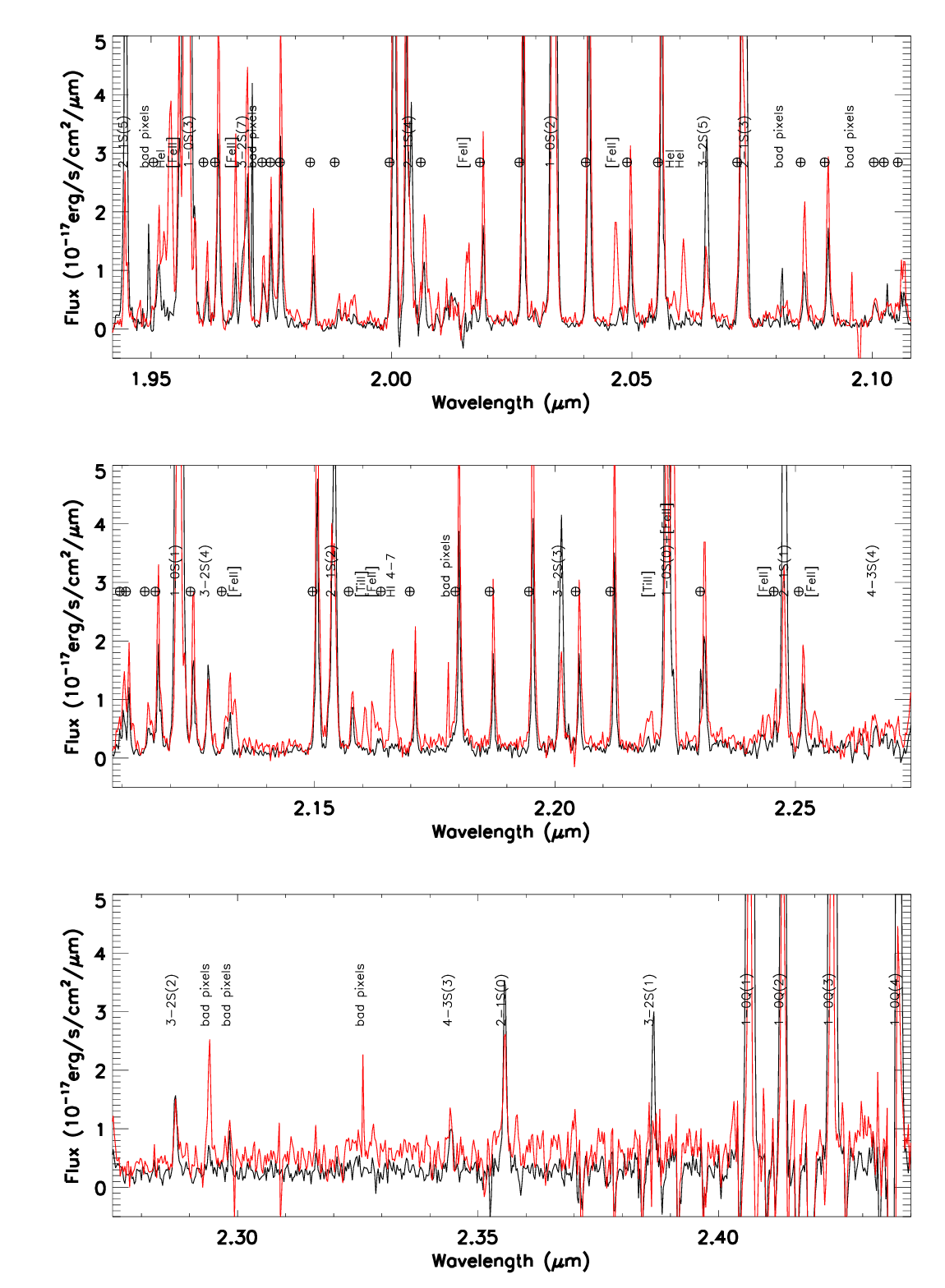}
\caption{\label{fig:spettro_k} As in Figure\, \ref{fig:spettro_j} for the K band.}
\end{figure*}

The SINFONI data reduction pipeline (Modigliani, Ballester \& Peron, 2007) has been used to subtract the sky
emission, to construct dark and bad pixel maps and flat field images, to correct for optical 
distorsions and to measure wavelength calibration by means of Xenon-Argon lamp images. 
Further analysis has been carried out with IRAF packages and
IDL scripts, that were used to remove telluric absorption features: this task was accomplished by dividing 
the target images by those of the telluric standard star, once corrected for both the stellar continuum
shape (a black-body function at the stellar temperature) and its intrinsic absorption features (mainly hydrogen
recombination lines). Photometric calibration was obtained  from the same standard star. Since there
is not overlapping in the spectral range covered by the different grisms, no cross-calibration has
been performed at this step of the data reduction. However, we have checked {\it a posteriori} 
the reliability of our absolute fluxes once an extinction map for the H$_2$ line emission has been obtained
(Sect.3.1.2): we have measured the departure from the theoretical value in the
image ratio of de-reddened lines coming from the same level and lying in different spectral bands. In this way 
we estimate a cross-calibration better than 5\% between J and H bands, and than 12\% between 
H and K bands.
Atmospheric OH lines were also used to refine wavelength calibration and to measure the effective spectral
resolution: we obtain {\it R}~$\sim$ 1900, 2700, 3500 in J, H, and K bands, that correspond to 160, 110, 85 km
s$^{-1}$, respectively. At these spectral resolutions we are able to resolve the brightest lines 
(e.g. H$_2$~2.122$\mu$m and [\fe]~1.644$\mu$m), which are observed at S/N ratios of $\sim$ 10$^2$-10$^3$.
As a final step, images of lines observed in individual bands ([\fe]1.257$\mu$m and 1.644$\mu$m in
the J and H bands and H$_2$ 2.122$\mu$m in the K band) were used to re-align the images acquired in 
different observing runs.\\
As a result, we obtained a 3D data-cube containing the HH99B image in more than
170 lines.\footnote{The reduced images are available at {\it http://cdsweb.u-strasbg.fr}.} As an example of our results, we have integrated the signal in the areas
corresponding to knots B0 and B3 (see Figure\,\ref{fig:righe}): the corresponding spectra are
shown in Figures\,\ref{fig:spettro_j}-\ref{fig:spettro_k}.
The large majority of the detected lines are H$_2$ ro-vibrational lines (121). For these, we list
in  Table~\ref{tab:tab1} spectral identification, vacuum wavelength, excitation 
energy (in K) and the maximum S/N ratio registered in the corresponding image. The detected ro-vibrational transitions 
come from levels with $v$ $\le$ 7 and E$_{up}$ up to $\sim$ 38~000 K, many of them never observed 
before in HH objects. In particular, as we show in Figure~\ref{fig:righe} (upper panel),
emission of lines with E$_{up}$ $\la$ 30~000 K is present only
along the bow flanks, while lines with E$_{up}$ $\ga$ 30~000 K are observed in the whole
shock, peaking at the bow head. 
Therefore, two main results emerge: {\it (i)} molecular hydrogen survives also where ionic emission
is strong (see below) and, {\it (ii)} temperature gradients do exist along the shock, with the highest 
values reached at the bow head, where stronger excitation conditions are expected to occur.\\
Atomic lines are listed in Table~\ref{tab:tab2} and some examples of the observations are shown
in Figure~\ref{fig:righe}, middle and bottom panels. Plenty of [\fe] lines are detected (34
lines), emitted from levels with E$_{up}$ $\la$ 30~000 K. As for H$_2$, two groups of 
lines are identified: those with E$_{up}$ $\la$ 13~000 K, which come for the a$^4$D level, are observed
in the whole region, while those at higher excitation energy are emitted only at the bow head.
In this same area, emission of hydrogen and helium recombination lines (8 and 3 lines, 
respectively, see Figure~\ref{fig:righe}, bottom panel) along with fine structure lines 
of [\pii],[\coii], and [\tiii] are detected. Other fine structure lines commonly observed in Herbig-Haro 
objects (e.g. [\ci]  at 0.98$\mu$m, [\n] at 1.04$\mu$m, [\sii] at 1.03$\mu$m, Nisini et al., 2002) 
are not covered with SINFONI. In fact, the first wavelength in the J band is $\lambda$=1.10$\mu$m,
i.e. longer than that of other infrared spectrographs (e.g. ISAAC and SofI at ESO are able to observe
wavelengths longer than 0.98$\mu$m and 0.86$\mu$m, respectively). 

To check for possible line variability, the intensities of atomic and molecular lines 
observed with SINFONI were compared with the ones observed by MC04 in July
2002 (i.e. four years before our SINFONI observations). 
Synthetic slits corresponding to slit 1 and 2 of MC04 (PA=32.4$^{\circ}$ and 329.5$^{\circ}$, 
width 0.6$^{\prime\prime}$) were superimposed onto the SINFONI
images and the flux was integrated over the same regions as in that paper.
The main difference between SINFONI and ISAAC spectra is represented by the significant
larger number of lines observed with SINFONI (see Figures\,\ref{fig:spettro_j}-\ref{fig:spettro_k} and
Figure~3 and Table~1 of MC04). In particular, it is 
noticeable the detection of many lines at high excitation, both atomic and molecular, that
indicate excitation conditions stronger than those inferred in that paper (see Table\,\ref{tab:tab4}). 
A general increase in the intensities of lines observed with both the instruments is registered, but with differential behaviour 
among molecular and ionic lines, which are respectively 4 and 1.5 times brighter than four years
before. 
Such variability has been observed in proper motion studies, over time periods of a few years
typical of radiative cooling times in HH objects (e.g. Caratti o Garatti et al., 2008).

\begin{table*}
\caption{\label{tab:tab1} H$_2$ lines observed in HH99B {\it (to be continued)}.}  \small
\begin{center}
\begin{tabular}{llrr|llrr|llrr}
\hline\\[-5pt]
\multicolumn{12}{c}{H$_2$ lines} \\
\hline\\[-5pt]
Line id. & $\lambda_{vac}$        & E$_{up}$    & S/N$_{max}^a$& Line id. & $\lambda_{vac}$  & 
E$_{up}$   & S/N$_{max}$ & Line id. & $\lambda_{vac}$      & E$_{up}$ & S/N$_{max}$\\
         & ($\mu$m)                 &  (K)       &            &           &   ($\mu$m)& 
     (K)&              &          & ($\mu$m)  &  (K)& \\

\hline\\[-5pt]
\multicolumn{12}{c}{$v$=1} \\
\hline\\[-5pt]
1-0S(0)$^1$ & 2.2233 & 6471 &44& 1-0S(7) & 1.7480 & 12818 & 65      & 1-0S(21)$^3$ & 1.7195 & 38136&7\\   
1-0S(1)    & 2.1218 & 6951 & 650 & 1-0S(8) & 1.7147 & 14221 & 70  &  1-0S(23) & 1.7801 & 42122&4 \\
1-0S(2)    & 2.0338 & 7584 & 48  & 1-0S(9) & 1.6877 & 15723 & 46  &  1-0Q(1) & 2.4066 & 6149&44   \\
1-0S(3)    & 1.9576 & 8365 & 46  & 1-0S(10)& 1.6665 & 17312 &  9  &  1-0Q(2) & 2.4134 & 6471&30   \\
1-0S(5)$^2$ & 1.8358 & 10342 & 5  &1-0S(18)$^3$& 1.6586 & 32136 & 3  &1-0Q(3) & 2.4237 & 6951&50     \\
1-0S(6) & 1.7879 & 11522    &  30&1-0S(19)$^3$ & 1.6750 & 34131 & 8&  1-0Q(4) & 2.4375 & 7584&18   \\
\hline\\[-5pt]
\multicolumn{12}{c}{$v$=2} \\
\hline\\[-5pt]
2-0S(0) & 1.1896 & 12095 & 6 	  &2-0Q(8) & 1.3020 & 16881     & 9     & 2-1S(1) & 2.2477 & 12551 & 44\\
2-0S(1) & 1.1622 & 12551 & 48	  &2-0Q(9)$^5$ & 1.3188 & 18108 &12     & 2-1S(2) & 2.1542 & 13151 & 22\\
2-0S(2) & 1.1382 & 13151 & 7 	  & 2-0Q(10) & 1.3375 & 19436 & 3      & 2-1S(3)& 2.0735 & 13891&40 \\
2-0S(3) & 1.1175 & 13891 & 6 	  & 2-0Q(15) & 1.4648 & 27267 & 3      & 2-1S(4) & 2.0041 & 14764& 21 \\
2-0S(4) & 1.0998 & 14764 & 6 	  & 2-0Q(19)$^3$ & 1.6156 & 34447 & 3      & 2-1S(5) & 1.9448  & 15764 &30 \\
2-0Q(1) & 1.2383 & 11790 & 14	  & 2-0O(2)  & 1.2932 & 11636 & 7      & 2-1S(8)$^6$ & 1.8183 & 19435&7 \\
2-0Q(2)$^4$ & 1.2419 & 12095&9	  & 2-0O(3)  & 1.3354 & 11790 & 16 	& 2-1S(9)$^7$  & 1.7904 & 20855&22 \\
2-0Q(3) & 1.2473 & 12551 &22 	 & 2-0O(6)  & 1.4870 & 13151 & 5      &2-1S(10)& 1.7688 &22356 & 3\\  
2-0Q(4) & 1.2545 & 13151 & 7 	   & 2-0O(7) & 1.5464 & 13891&8        & 2-1S(15)$^8$ & 1.7387 & 30794&6\\
2-0Q(5) & 1.2636 & 13891 & 11	  & 2-0O(9) & 1.6796 & 15764 & 5      & 2-1S(17)$^3$  & 1.7587 & 34446&4 \\
2-0Q(6) & 1.2745 & 14764 & 7  	   & 2-0O(11) & 1.8349 & 18108 & 3    & 2-1S(18)$^3$ & 1.7753 & 36301 & 3 \\
2-0Q(7) & 1.2873 & 15764 & 9         &2-1S(0) & 2.3556 & 12095 & 10    & 2-1S(19)$^{3,9}$& 1.7962 & 38166 & 8\\
   
\hline\\[-5pt]
\multicolumn{12}{c}{$v$=3} \\
\hline\\[-5pt]
3-1S(0)$^{10}$ & 1.2621 & 17388 & 9& 3-1S(9) & 1.1204  & 25661 & 22        & 3-1O(5) & 1.5220 & 17819& 10 \\
3-1S(1) & 1.2330 & 17819 & 10     & 3-1Q(1) & 1.3141 & 17099 & 8          & 3-1O(7)$^{14}$ & 1.6453 & 19087 & 9\\
3-1S(2) & 1.2077 & 18387 & 8      & 3-1Q(2)$^{12}$ & 1.3181 & 17388 & 12   & 3-1O(9)$^{11}$ & 1.7898 & 25661&10\\
3-1S(3) & 1.1857 & 19087 & 9      & 3-1Q(3) & 1.3240 & 17819 & 8          & 3-2S(1) & 2.3864 & 17819 & 8\\
3-1S(4) & 1.1672 & 19913 & 8      & 3-1Q(5)$^{13}$ & 1.3420 & 19087 & 9    & 3-2S(2) & 2.2870 & 18387 & 9 \\
3-1S(5) & 1.1520 & 20857 & 10     & 3-1Q(11)& 1.4479 & 25661 &  8	  & 3-2S(3) & 2.2014 & 19087& 25 \\
3-1S(6) & 1.1397 & 21912 & 5      & 3-1Q(13) & 1.5024 & 28558 & 6         & 3-2S(4) & 2.1280 & 19913 & 11 \\
3-1S(7) & 1.1304 & 23071 & 7      & 3-1Q(15)$^3$ & 1.5685 & 31691 &6 	   & 3-2S(5) & 2.0656 & 20857& 17 \\
3-1S(8) & 1.1241 & 24323 & 5      & 3-1O(4) & 1.4677 & 17388 &6           & 3-2S(7) & 1.9692 & 23071 & 14 \\
\hline\\[-5pt]
\end{tabular}
\end{center}
\end{table*}
\addtocounter{table}{-1}
\begin{table*}
\caption{\label{tab:tab1} H$_2$ lines observed in HH99B {\it (continued)}.}  \small
\begin{center}
\begin{tabular}{llrr|llrr|llrr}
\hline\\[-5pt]
\multicolumn{12}{c}{H$_2$ lines} \\
\hline\\[-5pt]
Line id. & $\lambda_{vac}$        & E$_{up}$    & S/N$_{max}^a$& Line id. & $\lambda_{vac}$  & 
E$_{up}$   & S/N$_{max}$ & Line id. & $\lambda_{vac}$      & E$_{up}$ & S/N$_{max}$\\
         & ($\mu$m)                 &  (K)       &            &           &   ($\mu$m)& 
     (K)&              &          & ($\mu$m)  &  (K)& \\
\hline\\[-5pt]
\multicolumn{12}{c}{$v$=4} \\
\hline\\[-5pt]
4-2S(0)$^{15}$ & 1.3425 & 22354 &9 & 4-2S(8) & 1.1987 & 28885 & 7& 4-2O(4)$^{18}$ & 1.5635 & 22354 & 6 \\
4-2S(1) & 1.3116 & 22760  & 14   & 4-2S(9) & 1.1958 & 30141 & 6	& 4-2O(5) & 1.6223 & 22760 & 7 \\
4-2S(2) & 1.2846 & 23296  & 21   & 4-2Q(7) & 1.4592 & 25625 & 6 & 4-2O(6) & 1.6865 & 23297 & 4  \\
4-2S(3)$^{16}$ & 1.2615 & 23956 &9 & 4-2Q(9) & 1.4989 & 27708 & 7 &4-2O(7) & 1.7563 & 23956 & 7\\
4-2S(4)$^{17}$ & 1.2422 & 24735 &9& 4-2Q(11)$^3$ & 1.5495 & 30141 & 3 &4-3S(3) & 2.3445 & 23956& 6 \\
4-2S(5) & 1.2263 & 25625 &12    & 4-2Q(13)$^3$ & 1.6123 & 32857 & 6 &4-3S(4) & 2.2667 & 24735& 3 \\
4-2S(6) & 1.2139 & 26618 & 5    & 4-2Q(15)$^3$ & 1.6892 & 35786 & 6 \\
4-2S(7) & 1.2047 & 27708 & 10    & 4-2O(3) & 1.5099 & 22081 & 6    \\
\hline\\[-5pt]
\multicolumn{12}{c}{$v$=5} \\
\hline\\[-5pt]
5-3S(3) & 1.3472 & 28500 &4    &  5-3Q(1) & 1.4929 & 26737 & 3 &  5-3Q(7)$^{3,19}$ & 1.5626 & 30065 & 6  \\
5-3S(4) & 1.3270 & 29231 & 3    &  5-3Q(2) & 1.4980 & 26994 & 9 & 5-3Q(11)$^3$& 1.6673 & 34291 & 4\\
5-3S(5) & 1.3107$^3$ & 30066 &4 & 5-3Q(4) & 1.5158 & 27880 & 5 & 5-3Q(12)$^3$ & 1.7021 & 35529 & 4 \\
5-3S(7) & 1.2894$^3$ & 32017 & 7    & 5-3Q(6) & 1.5443 & 29230& 3  & 5-3Q(13)$^3$ & 1.7412 & 36821 & 5\\

\hline\\[-5pt]
\multicolumn{12}{c}{$v$=6} \\
\hline\\[-5pt]
6-4S(1)$^3$ & 1.5015 & 31664 &  3 & 6-4Q(7)$^3$      & 1.6829 & 34175 & 5  & 6-4O(4)$^{3,21}$ & 1.7965 & 31306&8\\
6-4Q(3)$^3$    & 1.6162 & 31664 & 5 & 6-4Q(9)$^{3,20}$ & 1.7369 & 35992&6  \\
  \hline\\[-5pt]
\multicolumn{12}{c}{$v$=7} \\
\hline\\[-5pt]
7-5Q(5)$^3$ & 1.7784  & 36591 &3 &&&& \\
\hline\\[-5pt]  
\end{tabular}
\end{center}
Notes: $^a$maximum signal-to-noise ratio in the line image. In case of blends the reported number
refers to the sum of the blended lines (unless the emission comes from different zones of the bow). 
\begin{tabbing}
AAAAAAAAAAAAAAAAAAAAAAAAAAA \=  AAAAAAAAAAAAAAAAAAAAAAAAAAA   \= \kill
 $^1$ blends with ~[{\ion{Fe}{ii}}]\,a$^2\!$H$_{11/2}$-a$^2\!$G$_{9/2}$  \> $^8$ blends with 6-4Q(9), HI4-10    \>$^{15}$ blends with 3-1Q(5) \\
 $^2$ contaminated by atmospheric absorption                      \> $^{9}$ blends with 6-4O(4)	       \> $^{16}$ blends with 3-1S(0) \\
 $^3$ detected in the whole bow                                   \> $^{10}$ blends with 4-2S(3)        \>$^{17}$ blends with 2-0Q(2) \\
 $^4$ blends with 4-2S(4)                                         \>$^{11}$ blends with 2-1S(9)	        \>$^{18}$ blends with 5-3Q(7)\\
 $^5$ blends with 3-1Q(2)                                         \> $^{12}$ blends with 2-0Q(9)       \> $^{19}$ blends with 4-2O(4)\\
 $^6$ blends with HI 4-9                                         \> $^{13}$ blends with 4-2S(0)        \> $^{20}$ blends with 2-1S(15), HI 4-10\\
 $^7$ blends with 3-1O(9)                                        \> $^{14}$ blends with~[{\ion{Fe}{ii}}]\,a$^4\!$D$_{7/2}-$a$^4\!$F$_{9/2}$ \> $^{21}$ blends with 2-1S(19)\\
\end{tabbing}	
\end{table*}

\begin{table*}
\caption{\label{tab:tab2} Ionic lines observed in HH99B. }  \small
\begin{center}
\begin{tabular}{llrr|llrr}
\hline\\[-5pt]
\multicolumn{8}{c}{Ionic lines} \\
\hline\\[-5pt]
Line id. & $\lambda_{vac}$   & E$_{up}$   & S/N$_{max}$$^a$& Line id. & $\lambda_{vac}$  & E$_{up}$&S/N$_{max}$ \\
         &  ($\mu$m)          &  (K)       &               &        &  ($\mu$m)        &  (K)   &  \\
\hline\\[-5pt]
\multicolumn{8}{c}{[FeII] lines$^b$} \\
\hline\\[-5pt]
$\mathbf{a^4\!D_{7/2}}-$a$^6\!$D$_{9/2}$ & 1.2570 & 11446 & 700    &$\mathbf{a^4\!D_{1/2}}-$a$^6\!$D$_{3/2}$ & 1.2525 & 12729&20 \\
a$^4\!$D$_{7/2}-$a$^6\!$D$_{7/2}$ & 1.3209         & 11446 & 130    & a$^4\!$D$_{1/2}-$a$^6\!$D$_{1/2}$ & 1.2707 & 12729 & 24\\
a$^4\!$D$_{7/2}-$a$^4\!$F$_{5/2}$ & 1.9541         & 11446 &  7     & a$^4\!$D$_{1/2}-$a$^4\!$F$_{5/2}$ & 1.6642 & 12729 & 18 \\
a$^4\!$D$_{7/2}-$a$^4\!$F$_{9/2}$ $^1$ & 1.6440 & 11446     & 720    & a$^4\!$D$_{1/2}-$a$^4\!$F$_{3/2}$ & 1.7454 & 12729 & 21\\
a$^4\!$D$_{7/2}-$a$^4\!$F$_{7/2}$ & 1.8099        & 11446 &   99    & $\mathbf{a^4\!P_{5/2}}-$a$^4\!$D$_{5/2}$ & 1.9675 & 19387&20 \\
$\mathbf{a^4\!D_{5/2}}-$a$^6\!$D$_{9/2}$ & 1.1916 & 12074 & 7      & a$^4\!$P$_{5/2}-$a$^4\!$D$_{1/2}$ $^2$ &2.1609 &19387& 5\\  
a$^4\!$D$_{5/2}-$a$^6\!$D$_{7/2}$ & 1.2489         & 12074 & 17     & $\mathbf{a^4\!P_{3/2}}-$a$^4\!$D$_{7/2}$ & 1.7489 & 19673&9 \\ 
a$^4\!$D$_{5/2}-$a$^6\!$D$_{5/2}$ & 1.2946         & 12074 & 70     & $\mathbf{a^4\!P_{1/2}}-$a$^4\!$D$_{5/2}$ & 1.8139 &20006&62 \\ 
a$^4\!$D$_{5/2}-$a$^6\!$D$_{3/2}$ & 1.3281         & 12074 & 42     & $\mathbf{a^2\!G_{9/2}}-$a$^4\!$D$_{7/2}$ $^3$&1.2675&22797&7 \\
a$^4\!$D$_{5/2}-$a$^4\!$F$_{9/2}$ & 1.5339         & 12074 & 30     & $\mathbf{a^2\!G_{7/2}}-$a$^4\!$D$_{7/2}$ $^3$& 1.1885 &23552&16 \\
a$^4\!$D$_{5/2}-$a$^4\!$F$_{7/2}$ & 1.6773        & 12074 & 157     & $\mathbf{a^2\!P_{3/2}}-$a$^4\!$P$_{5/2}$ & 2.0466 &26417&4\\ 
a$^4\!$D$_{5/2}-$a$^4\!$F$_{5/2}$ & 1.8005 &     12074 &    35      & a$^2\!$P$_{3/2}-$a$^4\!$P$_{3/2}$ & 2.1334 & 26417 &5\\
$\mathbf{a^4\!D_{3/2}}-$a$^6\!$D$_{3/2}$ & 1.2791 & 12489& 41       & a$^2\!$P$_{3/2}-$a$^4\!$P$_{1/2}$ & 2.2442 & 26417 &4\\
a$^4\!$D$_{3/2}-$a$^6\!$D$_{1/2}$ & 1.2981 & 12489 & 23              & $\mathbf{a^2\!H_{11/2}}-$a$^2\!$G$_{9/2}$$^4$ & 2.2244&29265 & 21\\
a$^4\!$D$_{3/2}-$a$^4\!$F$_{7/2}$ & 1.5999 & 12489 & 46	            & $\mathbf{a^2\!H_{9/2}}-$a$^2\!$G$_{9/2}$ & 2.0157 & 29934 & 10\\
a$^4\!$D$_{3/2}-$a$^4\!$F$_{5/2}$ & 1.7116& 12489 & 13		    & a$^2\!$H$_{9/2}-$a$^2\!$G$_{7/2}$ & 2.2541 &29934 &5\\ 
a$^4\!$D$_{3/2}-$a$^4\!$F$_{3/2}$ & 1.7976 & 12489 & 36 	    & $\mathbf{b^4\!P_{1/2}}-$a$^4\!$P$_{3/2}$ & 1.1446&32242&4\\ 

\hline\\[-5pt]						 				
\multicolumn{8}{c}{H lines} \\
\hline\\[-5pt]
3-5 (Pa$\beta$) & 1.2822 & 151492 & 21        & 4-11 & 1.6811 & 156499 & 6 \\
4-14 & 1.5884  & 156999 &  3                  & 4-10$^5$ & 1.7367 & 156226 &6 \\
4-13 & 1.6114 & 156870 & 4                   & 4-9$^6$ & 1.8179 & 155855 & 6 \\ 
4-12 & 1.6412 & 156708 & 6                   & 4-7 (Br$\gamma$)& 2.1661 & 154583& 12\\    
\hline\\[-5pt]
\multicolumn{8}{c}{Other lines} \\
\hline\\[-5pt]
~{\ion{He}{i}}\,$^1P_{1}-^1\!S_{0}$ &2.0587  & 246226& 3 &              ~[{\ion{Ti}{ii}}]a$^2F_{5/2}$-a$^4F_{3/2}$$^{10}$ &2.1605 & 6652& 5\\ 
~{\ion{He}{i}}\,$^3D-^3\!P$ $^7$     & 2.0607 & 282101&8 &              ~[{\ion{Ti}{ii}}]a$^2F_{7/2}$-a$^4F_{9/2}$ &2.2201 & 7040& 5\\ 	
~{\ion{He}{i}}\,$^3P_{0}-^3\!D$ $^7$ & 1.9522 & 289992& 5 &              ~[{\ion{Ti}{ii}}]a$^2D_{5/2}$-a$^4F_{5/2}$ &1.1560 & 12570& 3\\ 
~[{\ion{P}{ii}}]$^2D_2$-$^3P_1$  &1.1471  & 12764 & 5    &		 ~[{\ion{Ti}{ii}}]a$^4P_{3/2}$-b$^4F_{7/2}$ &1.2036 & 13506& 3\\ 
~[{\ion{P}{ii}}]$^2D_2$-$^3P_2$ $^8$&1.1886  & 12764 &16     &  	    ~[{\ion{Ti}{ii}}]a$^2P_{3/2}$-b$^4F_{3/2}$ &1.1028 & 14340&3 \\ 
~[{\ion{Co}{ii}}]\,$b^3\!F_{4}$-a$^5\!F_{5}$ $^9$ & 1.5474 & 14119& 5   &   ~[{\ion{Ti}{ii}}]a$^2H_{11/2}$-a$^2F_{/2}$ &1.2695 & 17729& 5\\ 

\hline\\[-5pt]

\end{tabular}
\end{center}
Notes: $^a$maximum signal-to-noise ratio in the line image. In case of blends the reported number
refers to the sum of the blended lines (unless the emission comes from different zones of the bow). 
$^b$ Lines coming from the same upper level are grouped, and the first term of each group is 
evidenced with bold-face characters.
\begin{tabbing}
AAAAAAAAAAAAAAAAAAAAAAAAAAAAAAAAAAAAAAAAAAAA   \= \kill
$^1$ blends with 3-1O(7)  \> $^6$ blends with 2-1S(8)\\
$^2$ blends with [{\ion{Ti}{ii}}]a$^2F_{5/2}$-a$^4F_{3/2}$ (fundamental transition) \> $^7$ multiplet\\
$^3$ blends with [\pii]$^1D_2$-$^3P_2$  \> $^8$ blends with [\fe] a$^2\!G_{7/2}$-a$^4\!$D$_{7/2}$\\
$^4$ blends with 1-0S(0)  \>  $^9$ tentative identification \\
$^5$ blends with 2-1S(15), 6-4Q(9) \> $^{10}$ blends with [\fe] a$^4\!$P$_{5/2}-$a$^4\!$D$_{1/2}$\\
\end{tabbing}
\end{table*}

\section{Diagnostics of physical parameters}

\subsection{Fe analysis}

\subsubsection{On [\fe] Einstein coefficients}
In Table\,\ref{tab:tab2} the [\fe] lines are listed grouping transitions originating
from the same upper level. The intensity ratio of pairs of (optically thin) lines in each group 
is independent of the local physical conditions, being function only of atomic parameters (i.e. line frequencies and 
Einstein {\it A} coefficients for spontaneous emission). Hence, the {\it observed} intensity ratios 
among these lines can be efficiently used to measure the local extinction (e.g. Gredel, 1994). 
Unfortunalety, the complexity of the energy level system of iron makes very difficult to
accurately compute the {\it A} values, so that three distinct sets of these
parameters, differing by more than 30\%, have been sofar listed 
in the literature: two were computed with different methods by Quinet, Le Dourneuf
\& Zeippen, (Q-SST, Q-HFR, 1996), and one is provided by Nussbaumer \& Storey (NS, 1988). A fourth list,
based on the observation of P\,Cygni, was recently published by Smith \& Hartigan 
(SH, 2006) who find {\it A} values 10\%-40\% higher than theoretical computations.
The application of one set rather than another can lead to significant differences in the
derivation of the local extinction. For example, the NS coefficients of the 1.644$\mu$m and 
1.257$\mu$m lines (both coming from the level a$^4D_{7/2}$ with E$_{up}$=11446 K and more commonly observed) provide an extinction 2.7~mag
higher than that derivable from the Q-SST coefficients. This marginally affects the NIR lines 
(the intensity grows by a factor of 2.8 for a line at 1$\mu$m and a by factor of 1.4 for a line at
2$\mu$m), but becomes critical at optical wavelengths (the intensity grows by a factor of 33.6
at 0.5$\mu$m).\\
The large number of [{\fe}] lines detected in HH99B, observed with a 
high S/N ratio 
in a remarkably large fraction of pixels, offers us the opportunity
to compare theoretical predictions on the spontaneous emission rates with a 
significant sample of observational points. 

\begin{figure*}
\centering
\includegraphics[width=12cm]{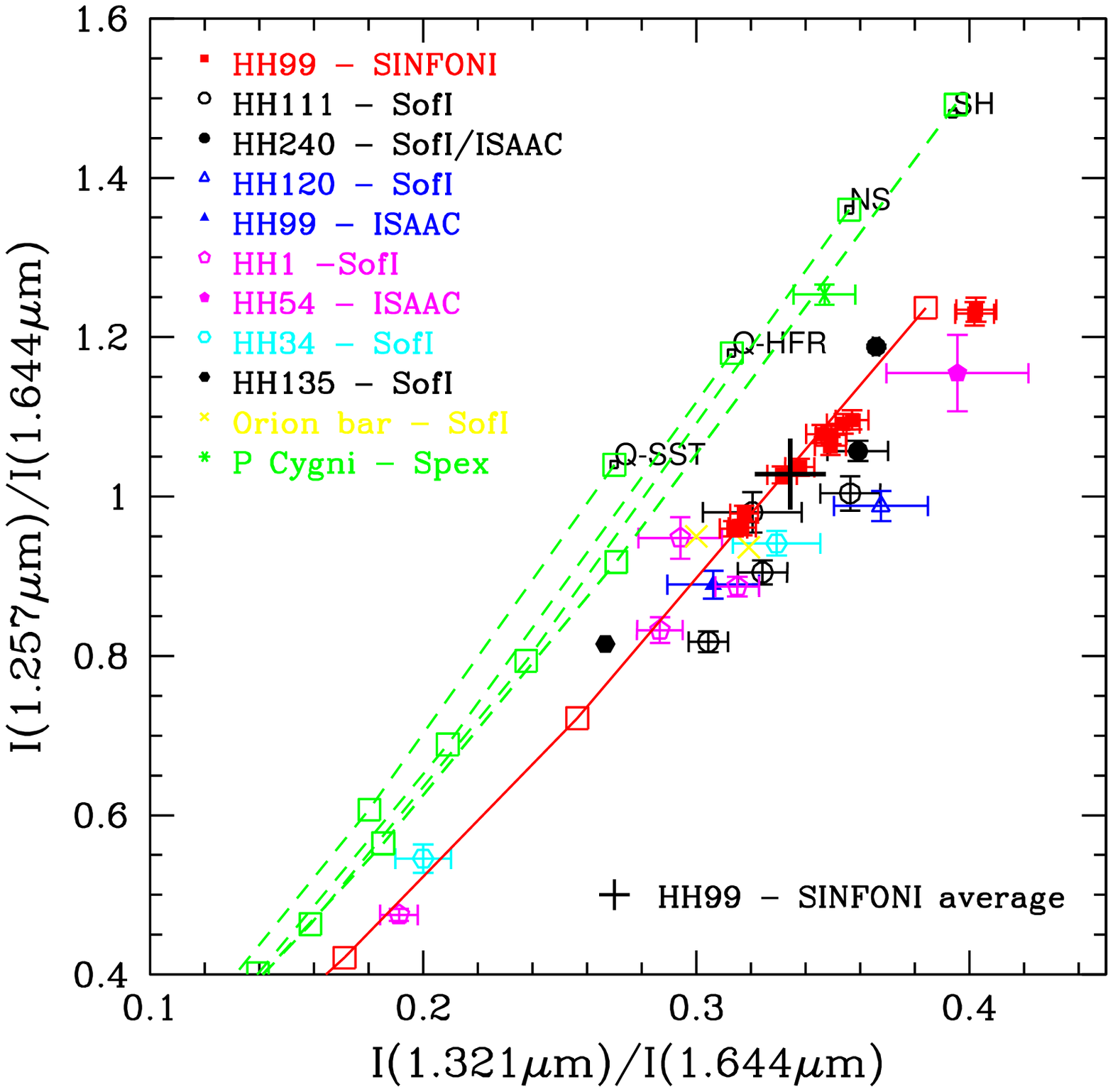}
\caption{\label{fig:av}I(1.257$\mu$m)/I(1.644$\mu$m) vs. I(1.321$\mu$m)/I(1.644$\mu$m) 
ratios measured in different objects (depicted with different
symbols/colors). The HH99B-SINFONI data (red squares) have been 
computed in pixels where the S/N of each of the three lines
is larger than 100, while other data are literature observations where S/N $\ge$ 30. Intrinsic line ratios predicted
theoretically (Q-SST = Quinet et al. 1996 - SuperStructure; Q-HFR =
Quinet et al., 1996 - Relativistic Hartree-Fock; NS = Nussbaumer \& Storey, 1988),
along with the observational point  (SH) by Smith \& Hartigan (2006),
are labelled. Green dashed curves represent the extinction law
by Rieke \& Lebofsky (1985), starting from different theoretical points; open squares refer to
A$_V$\,=\,0\,,5\,,10 mag. The same extinction law (in red) has been applied to the A$_V$ = 0 mag
point derived from SINFONI data. This latter has been derived by applying to the
average of the HH99B points (black cross) a visual extinction of 1.8\,mag, as 
estimated  from the P$\alpha$/Br$\gamma$ ratio (see text). References: HH99B - SINFONI: this work;
HH111-, HH240-, HH120-SofI: Nisini et al., 2002; HH240- ISAAC: Calzoletti et
al., 2008; HH99B - ISAAC : M$^c$Coey et al., 2004; HH1 - SofI: Nisini 
et al., 2005; HH54 - ISAAC: Giannini et al., 2007; HH34 - SofI: Podio et al.,
2006; HH135 - SofI: Gredel, 2006; Orion bar - SofI: Walmsley et al., 2000; P
Cygni - Spex: Smith \& Hartigan, 2006.}
\end{figure*}

To this aim, we have plotted in Figure\,\ref{fig:av} 
the ratios I(1.257$\mu$m)/I(1.644$\mu$m) vs. I(1.321$\mu$m)/I(1.644$\mu$m), since these are observed 
at very high S/N (larger than 100, red filled squares). 
In the same Figure, green dashed curves represent the Rieke \& Lebofsky (1985) extinction 
law\footnote {The following arguments remain valid if other extinction laws are adopted (e.g. Cardelli, 
Clayton \& Mathis, 1988), since these do not appreciably differ from the Rieke \& Lebofsky law over 
the short wavelength range taken into consideration (from 1.257$\mu$m to 1.644 $\mu$m).} applied to the
intrinsic ratios expected for the four sets of {\it A} coefficients (Q-SST, Q-HFR, NS, and SH).
Squares along these 'extinction curves' indicate A$_V$\,=\,0,~5,~10\,mag. 

First of all, we note that all the HH99B data lie definitively to the right of any of the plotted
extinction curves. Since different A$_V$ can move the points only {\it along} extinction vectors, no 
theoretical intrinsic ratio is consistent with the observed points. This result was already pointed
out in Nisini et al. (2005), who discussed how the extinction along the knots of the HH1 
jet determined from the 1.321$\mu$m/1.644$\mu$m ratio is always smaller than that derived from the
1.257$\mu$m/1.644$\mu$m ratio, irrespective of the adopted theoretical set of {\it A} coefficients.    
HH99B data are also inconsistent with the P\,Cygni datum (green triangle in
Figure\,~\ref{fig:av}), and consequently with the {\it A} coefficients extrapolated from it.\\ 
To be confident of the reliability of our data, we have checked for possible 
blendings of the [\fe] lines with other lines or telluric features. In this respect, 
the 1.644$\mu$m line is close to both the H$_2$ line 3-1O(7) at 1.6453$\mu$m (see Table\,\ref{tab:tab1} 
and Figure\,\ref{fig:spettro_h}) and a telluric OH feature at 1.6442$\mu$m 
(Lidman \& Cuby, 2000). This latter has been removed in the sky subtraction procedure, and its
residuals have been estimated to affect the 1.644$\mu$m flux by less than 0.5\%. 
The H$_2$ line, coming from a low excitation level (E$_{up}$ $\sim$ 19000 K), is observed only 
in the receding parts of the bow, and thus it results {\it spatially} separated from the 1.644$\mu$m emission
region.\\
To check for other observational or data-reduction biases (e.g. unfavourable observational conditions, 
flat-fielding, intercalibration of lines lying in different bands),
we searched in the literature for other observations of the considered lines obtained with other
instruments. To minimise the uncertainties, we have considered only line ratios observed with 
S/N $\ge$~30, that are shown with different colours/symbols in Figure\,~\ref{fig:av}. Notably, 
all of them occupy the right side of the plot, in agreement 
with the HH99B points. This result, that reinforces the reliability of our observations, 
allows us to derive new {\it A} coefficients from our observations, provided that these are
accurately corrected for the visual extinction value (measured independently from [\fe] lines).
In this respect, two facts have to be noted: {\it(i)} although the sky area considered for this analysis 
is a few arcsec$^2$ (marked in black in Figure~\ref{fig:righe}, middle left panel), an extinction gradient of $\sim$ 1 mag
occurs in this zone, as evidenced by the scatter among the data points (red squares) of 
Figure~\ref{fig:av}; {\it(ii)} in the same area, we are able to obtain just a gross estimate of A$_V$ 
(1.8$\pm$1.9 mag) from the observed P$\alpha$/Br$\gamma$ ratio (see Sect.~3.3). Both these
circumstances prevent us from deriving an accurate measure of the {\it A} ratios.
A rough estimate can be however obtained de-reddening   
the average of the HH99B data (black cross) for A$_V$\,=\,1.8~mag, which gives
A$_{1.321}$/A$_{1.644}$ = 0.38 and A$_{1.257}$/A$_{1.644}$ = 1.24. The main uncertainty
on these values comes, more than from the error on the line fluxes, from the
A$_V$ estimate. Taking into account the A$_V$ error of 1.9~mag, 
we can state that the theoretical points do belong to the segment of the extinction curve
(along which they are constrained to move) starting at the 
point [0.33,1.02] (A$_V$\,=\,0~mag) and ending at the point [0.44,1.48] (A$_V$\,=\,3.7~mag) .

In conclusion, from a pure observational point of view, we can summarise as follows: {\it (i)} all
the theoretically derived {\it A} values fail to reproduce the large majority of the observed line ratios,
irrespective of the extinction values; {\it (ii)} the best 'recipe' to derive a reliable 
extinction estimate from [\fe] lines is (at least when only the three considered lines are detected)
to use the NS coefficients for the 1.321$\mu$m/1.644$\mu$m ratio and the Q-HFR coefficients for the 
1.257$\mu$m/1.644$\mu$m ratio, that are 8\% and 5\% lower than our determinations; 
{\it (iii)} dedicated observations of objects with well known visual extinction should be
performed to derive the [\fe] Einstein {\it A} coefficients with sufficient accuracy.

\begin{figure*}
\centering
\includegraphics[width=12cm]{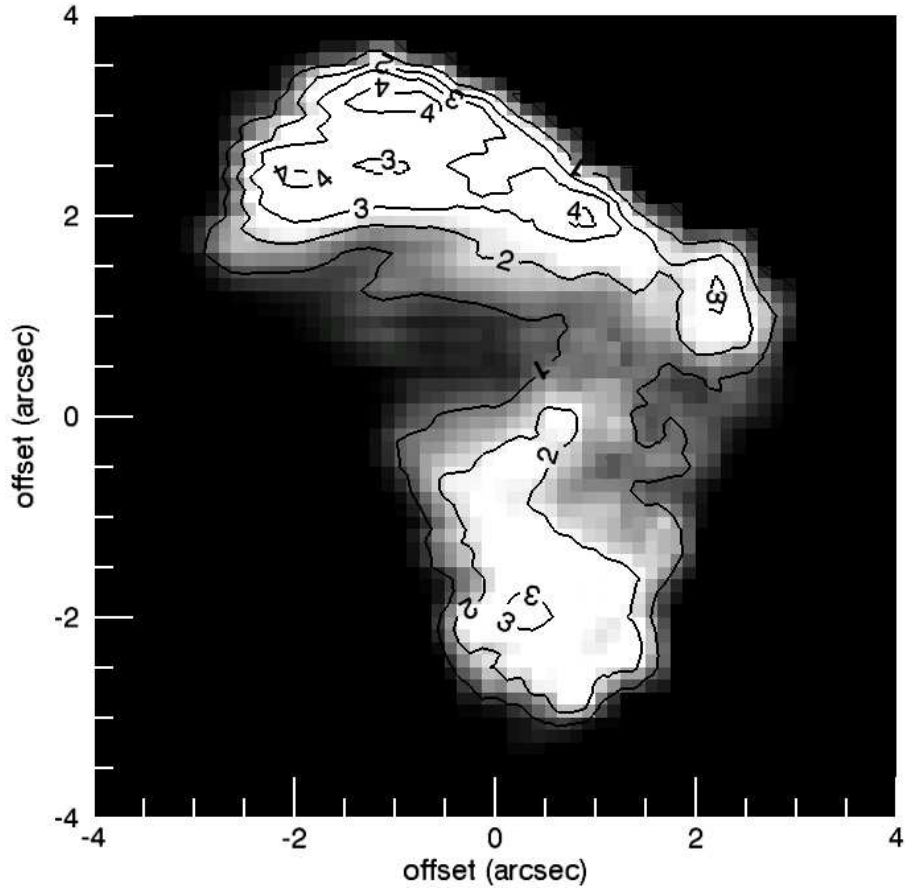}
\caption{\label{fig:av_fe} Extinction map obtained from [\fe] lines where contours from A$_V$=1~mag
to A$_V$=4~mag are shown. The NS Einstein {\it A} coefficients have been adopted.}
\end{figure*}
\begin{figure*}
\centering
\includegraphics[width=5cm]{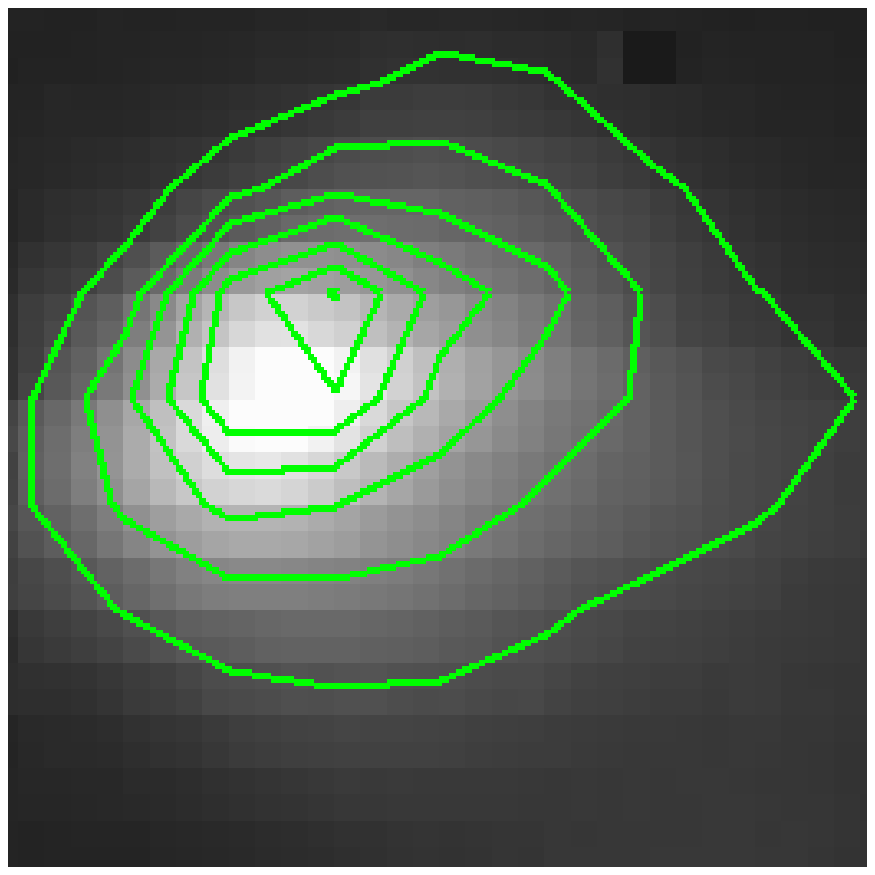}
\caption{\label{fig:intrinsic}. De-reddened contours of 1.257$\mu$m intensity, overlaied 
with the image acquired in the same line.}
\end{figure*}

\subsubsection{Extinction map}
Given the problems with the {\it A} coefficients outlined in the previous section, we
have applied the following procedure to construct an extinction map across the HH99B bow 
from the observed [\fe] lines: to minimise the effect of the uncertainties we used a number
of line ratios involving bright lines from four energy levels (i.e. a$^4$D$_{7/2}$, a$^4$D$_{5/2}$, 
a$^4$D$_{3/2}$, a$^4$P$_{5/2}$) and distant in wavelength. 
With this set of ratios, and adopting, as a first attempt the NS coefficients,
we determined the extinction in a very small region at the bow-head,
where all the lines are detected with S/N ratio larger than 30. This value has then been used
to calibrate the extinction map obtained from the 1.25$\mu$m/1.644$\mu$m ratio, which
is the only one detected well above the noise level (at least at 5$\sigma$) also in the bow flanks. 
Contours of the final map are shown in Figure\,\ref{fig:av_fe}: in a total area of $\sim$ 10\,arcsec$^2$
variations of A$_V$ up to 4 mag are recognised. The highest A$_V$ values (4-5\,mag) 
are found at the bow-head: if we correct the observed [\fe] lines for them, the emission
peak (see  Figure~\ref{fig:intrinsic}) moves of about 0.6 arcsec towards north-east.
Along the flanks A$_V$ is generally lower (up to 2-3~mag). Thus, the progressive fading 
in these zones of the \fe\, emission can not be ascribed to an increasing extinction, but 
rather reflects low excitation conditions and/or low abundance of the gas-phase iron.\\
As described above, the main uncertainty on the extinction map arises from the adopted set of the
{\it A} values.  We have thus re-derived the same map from the 1.257$\mu$m/1.644$\mu$m ratio, now 
adopting our Einstein coefficients ratio of 1.24. We find that the 
largest difference between the two maps occurs at the emission peak, where it is of $\sim$ 0.6 mag.
This implies a marginal increase in intrinsic line intensities (for example I(1.644$\mu$m) increases
by 10\%) and does not critically affects the derivation of the physical parameters of the atomic gas
(see next section). 

\subsubsection{Electron density map and temperature}

\begin{figure*}
\centering
\includegraphics[width=12cm]{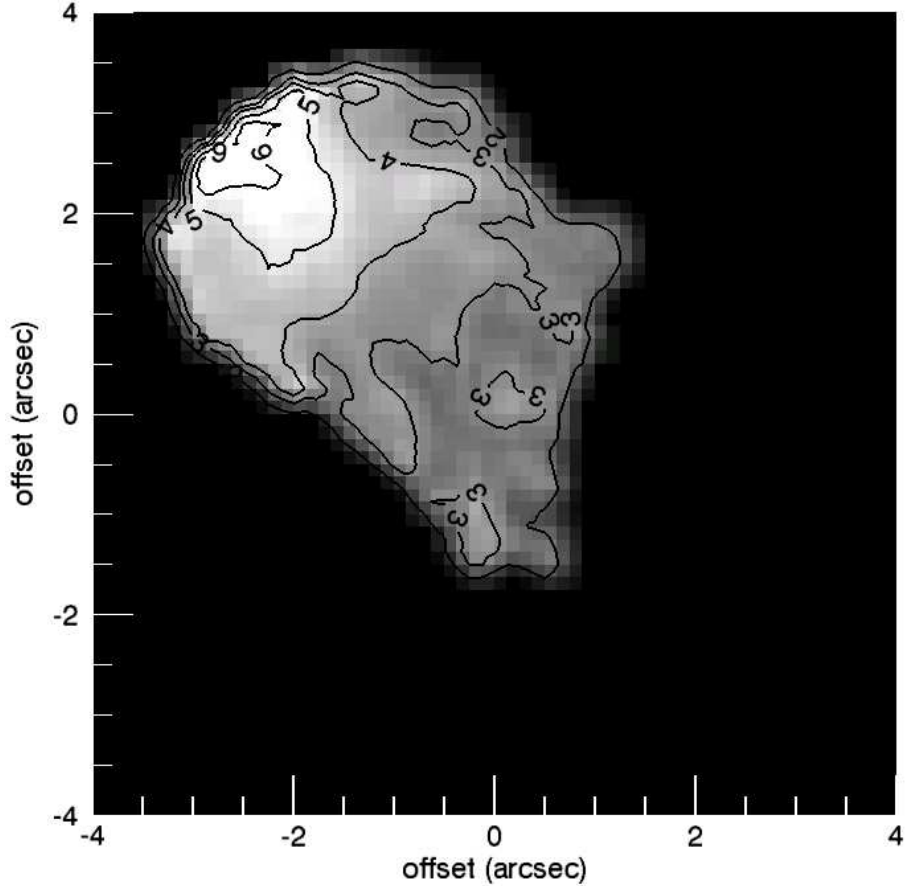}
\caption{\label{fig:den_fe} Electron density map as derived from [\fe] line ratios.
Contours are in units of 10$^3$ cm$^{-3}$.}
\end{figure*}

\begin{figure*}
\centering
\includegraphics[width=12cm]{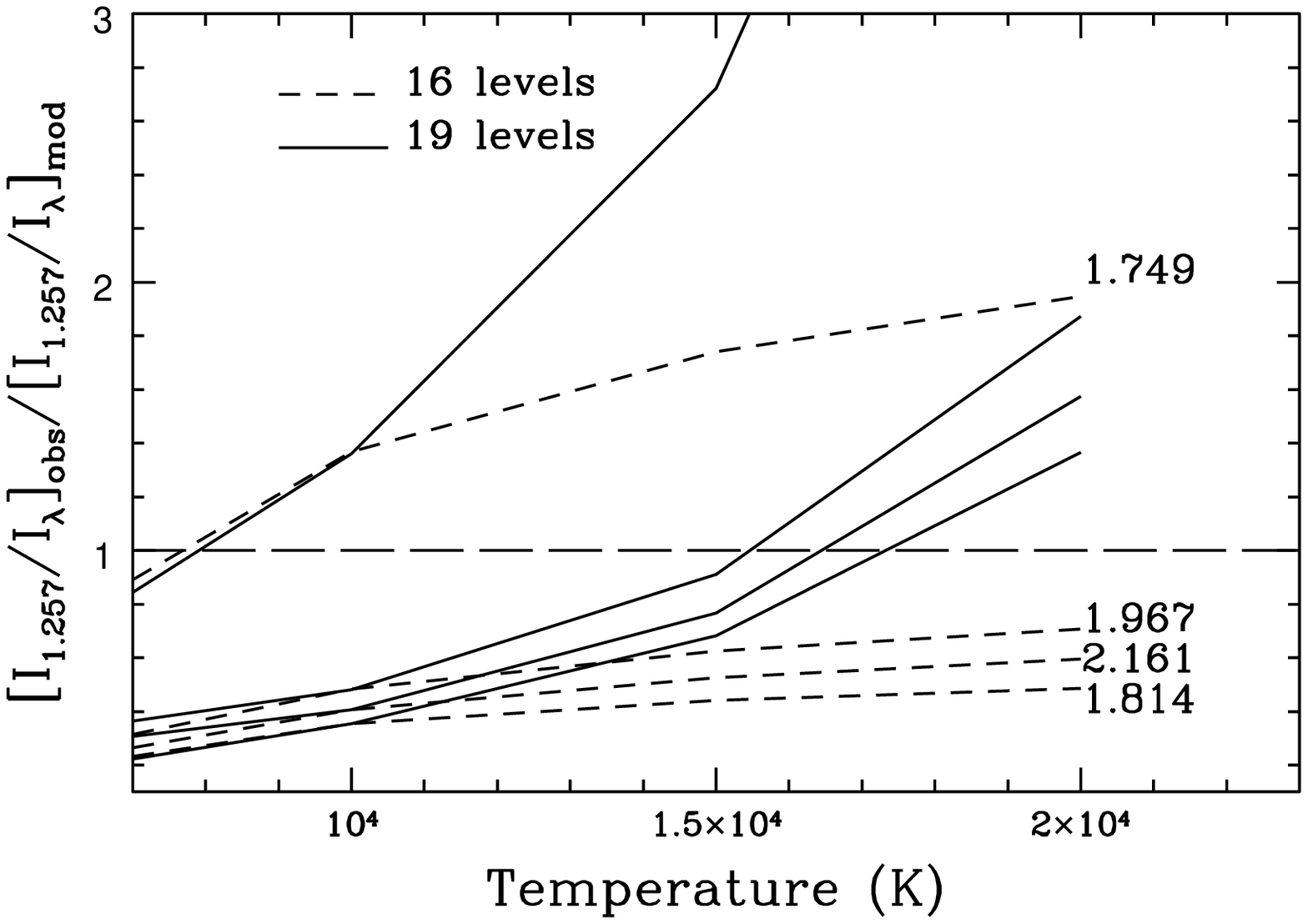}
\caption{\label{fig:temp_fe} [I$_{1.257}/I_{\lambda}]_{obs}/[I_{1.257}/I_{\lambda}]_{mod}$
plotted vs. the electronic temperature for four [\fe] lines. The results for the
\Fe\, NLTE model at 16 and 19 levels are shown for comparison.}
\end{figure*}

To derive the electron density along the bow structure, we selected seven intensity ratios 
(i.e. $I_{1.533}$/$I_{1.644}$, $I_{1.600}$/$I_{1.644}$, $I_{1.677}$/$I_{1.644}$, $I_{1.664}$/$I_{1.644}$, 
$I_{1.271}$/$I_{1.257}$, $I_{1.279}$/$I_{1.257}$, $I_{1.328}$/$I_{1.257}$), involving lines close in 
wavelength (their dfferential extinction is negligeable) and coming from levels with different 
critical densities (from $\sim$ 8~10$^2$ to 3~10$^5$ cm$^{-3}$) and similar excitation energy 
(E$_{up}$ $\sim$ 11\,000-12\,000 K), so that the dependence on the temperature is very weak. 
All these line ratios have been simultaneously fitted with a NLTE code which solves the equations
of the statistical equilibrium for the first 16 fine structure levels of [\fe]. Spontaneuos rates are
taken from NS, while energy levels and rates for electron collisions are adopted from Pradhan \& Zhang 
(1993). Assuming T$_e$~=~10\,000 K, we have constructed the electron density map shown in
Figure\,\ref{fig:den_fe}; n$_e$ is of the order of 2-4~10$^3$ cm$^{-3}$, with a peak up to 
6\,10$^3$ cm$^{-3}$ at the bow head\footnote{As we will see in the following, we estimate in
this part of the bow T$_e$$\sim$ 16000 K. At this temperature the ratios involved in the electron
density estimate can differ by less than 15\% with respect to those computed at T=\,10000 K. This
implies a very marginal increase of n$_e$.}. These values are in the range commonly found in HH
objects from embedded jets (e.g. Nisini et al., 2005, Podio et al., 2006).

In a restricted area at the bow head of about 1 arcsec$^2$ (see Figure\,\ref{fig:righe}, middle right
panel), we have detected 13 lines at high excitation (E$_{up}$ between 20~000 and 30~000 K), 
which are suitable for evaluating the local electronic temperature. Of these, 
just four lines coming from the term a$^4$P (i.e. 1.749$\mu$m, 1.814$\mu$m, 1.967$\mu$m, 2.161$\mu$m)
can be modelled, since for the remaining nine lines the collisional rates are unknown. 
 
To that aim, we have enlarged our 16 level code 
by including further three fine structure levels for which 
the collisional coefficients are reported by Zhang \& Pradhan (1995). 
Notably, the excitation energy of level \#19 is around 32~000 K, well above than 
that of level \#16 (less than 20~000 K): hence the temperature range that can be probed
with the 19 level code is sensitively enlarged. 
Having fixed extinction and electron density from the maps of Figures\,~\ref{fig:av_fe}
and~\ref{fig:den_fe}, we fitted the de-reddened ratios with the 1.257$\mu$m line, 
integrated over the area where the a$^4$P lines are detected at S/N $\ge$5. 
Results are plotted in Figure\,\ref{fig:temp_fe},
where the observed ratios are compared with the predictions of both
the 16 and 19 level codes and for temperatures from 10~000 to 20~000 K. 
First, we note that while at T$_e$$\la$10~000 K the inclusion of three further levels does not 
change the results of the 16 level code, strong differences emerge at higher temperatures 
(e.g. the ratio 1.257/1.749 decreases by about 70\% at T$_e$=20~000 K). Second, ratios with the 
1.814$\mu$m, 1.967$\mu$m, and 2.161$\mu$m lines
well agree with T$_e$$\sim$ 16~000 - 17~000 K (the latter line has been de-blended from the fundamental
line of [\tiii], using a NLTE model for this species, Garcia-Lopez et al., 2008). 

Finally, we note that the ratio with the 1.749$\mu$m line implies T$_e$$\sim$8~000 K, that we
consider not reliable because in the same spatial region examined here also bright hydrogen 
and helium recombination lines are observed (see Figure\,\ref{fig:righe}). 
For this line, however, neither evident discrepancies in the different computations of the Einstein 
coefficients (all the available lists give similar values, see Sect.3.1.1), nor observational
biases (e.g. extinction, blending with other lines) are able to explain the disagreement with 
the other ratios.

\subsubsection{[\fe] abundance}

\begin{figure*}
\centering
\includegraphics[width=12cm]{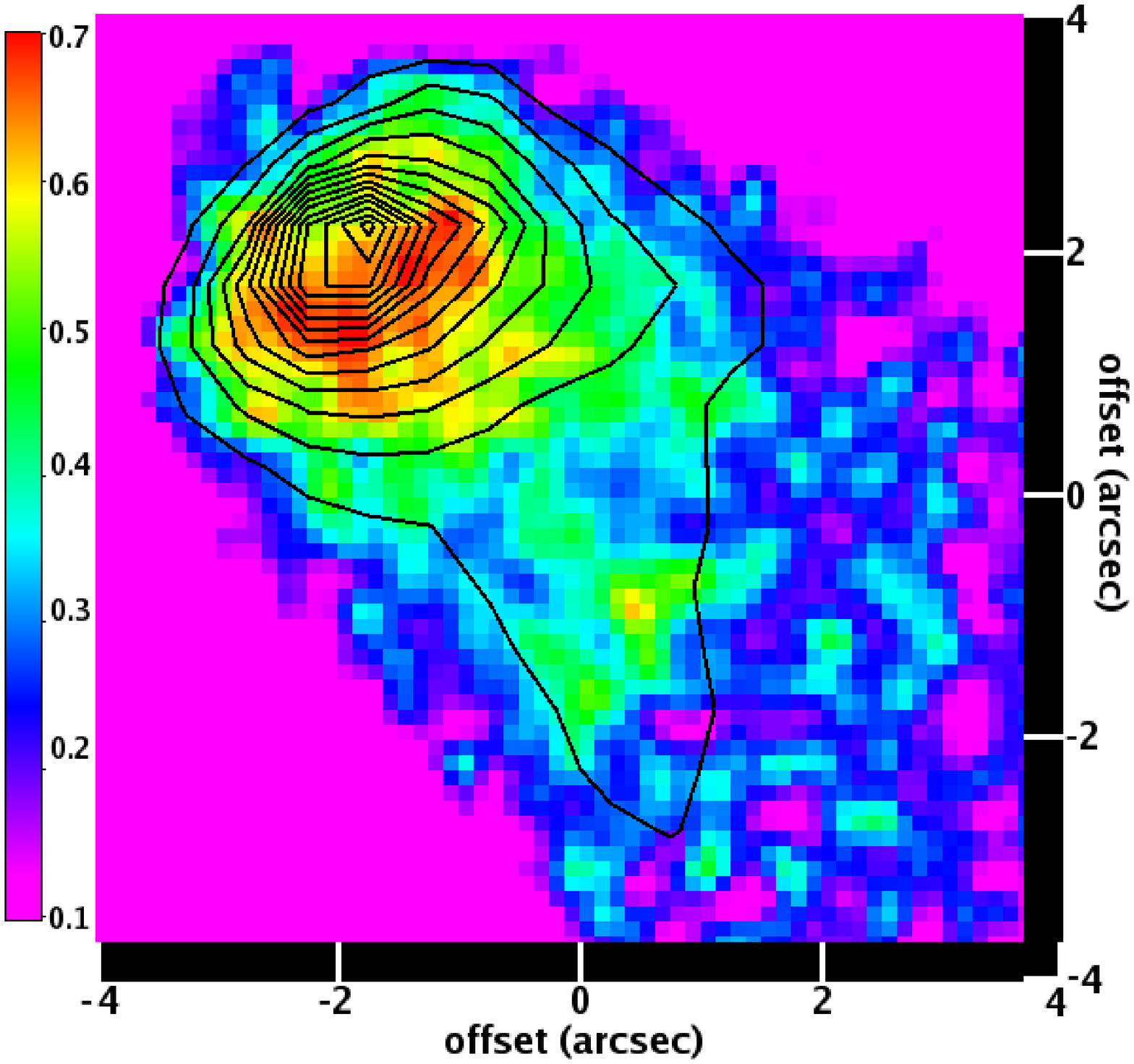}
\caption{\label{fig:abund_fe} Map of the percentage of iron in gas-phase with
overlaied the intensity contours of the [\fe] 1.257$\mu$m line.}
\end{figure*}

The gas-phase \Fe\, abundance x(\Fe) is a measure of the shock efficiency in disrupting the cores of the
dust grains where iron is locked in quiescent conditions (e.g. May et al., 2000). Estimates
of x(\Fe) in shock environments so far have given sparse results, from values
close to solar abundance (e.g. Beck-Winchatz, Bohm \& Noriega-Crespo, 1996), up to intermediate (Nisini et
al., 2002) and very high depletion factors (Mouri \& Taniguchi, 2000, Nisini et al., 2005).
A powerful way to estimate the percentage of gas-phase iron ($\delta_{Fe}$) 
based on [\fe]/[\pii] line ratios has been proposed by Oliva et al. (2001). Since phosphorus and iron
have similar ionisation potentials and radiative recombination coefficients, they are expected to be in the
first ionised state in comparable percentages; moreover, the near-IR lines of \fe\, and \pii\, 
have similar excitation energies and critical densities and therefore are
excited in similar physical conditions. Hence, [\fe]/[\pii] line ratios are good indicators of the 
relative abundance of the two species, and more specifically, because phosphorus is a non-refractory 
species, of the degree of iron depletion. 
In HH99B two [\pii] lines are detected, at 1.1471 and 
1.1885~$\mu$m, this latter barely blended with a [\fe] line (see Table~\ref{tab:tab1}). 
Oliva et al. derive: 

\begin{equation}
\frac{x(\Fe)}{x(\Po)}\simeq 2\times\frac{I([\fe]1.257)}{I([\pii]1.188)}\simeq\frac{I([\fe]1.257)}{I([\pii]1.147)} 
\end{equation}

This equation, as stated by the authors, is accurate to within a factor of 2 for all temperatures
and densities expected within the shock. By assuming a solar \Fe/\Po\, abundance ratio of $\sim$ 120
(Asplund, Grevesse \& Sauval, 2005) we have derived a map of the percentage of gas-phase
iron (see Figure\,\ref{fig:abund_fe}). A strong decrease in the percentage of gas-phase iron
occurs from the bow-head (70\%) towards the zones behind (up to 20\%), with an average
uncertainty of about 15\%, estimated from the propagation of the errors in both the two 
considered images. Notably, theoretical predictions for the degree of iron depletion as a
function of the shock velocity (Jones, 2000) imply that this latter should exceed
100 km s$^{-1}$ for $\delta_{Fe}$ $\ga$ 0.20, a condition verified in our case even in the bow flanks.


\subsection{H$_2$ analysis}
\subsubsection{Extinction map}

\begin{figure*}
\centering
\includegraphics[width=12cm]{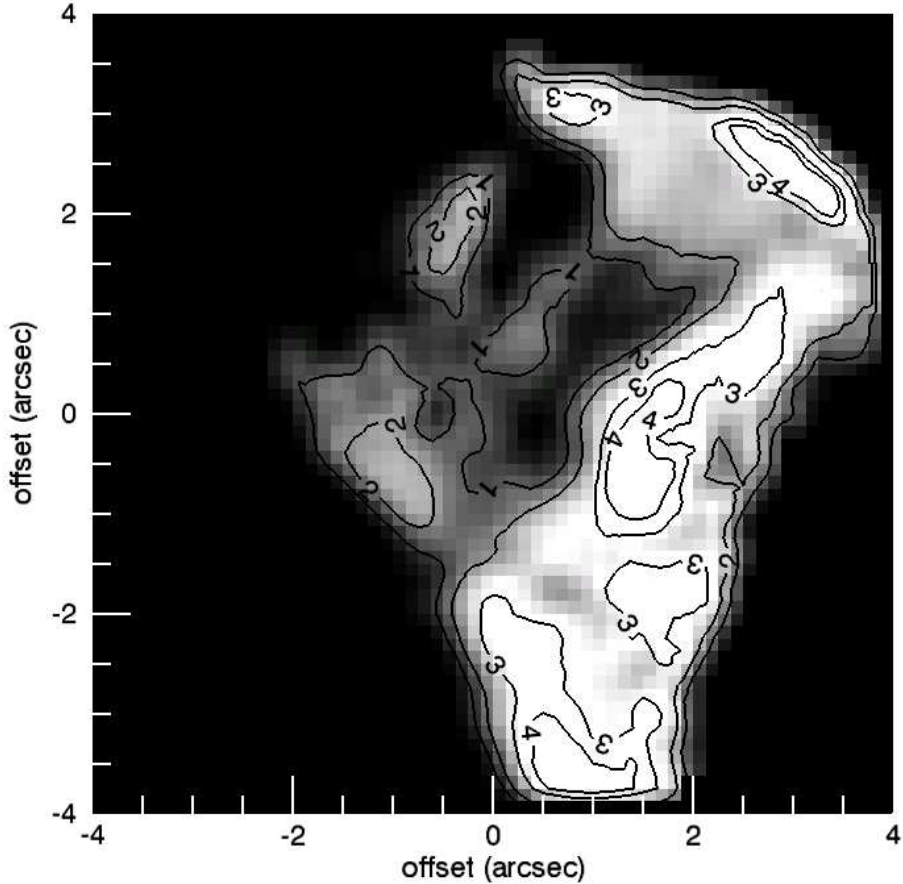}
\caption{\label{fig:av_h2} Extinction map as derived from H$_2$ line ratios.
Contours from A$_V$=1 to 4 mag are shown.}
\end{figure*}
Looking at the line images of Figure\,\ref{fig:righe} it is evident that the bulk of the
H$_2$ emission comes from the bow flanks. Since these parts of the bow are not covered in the 
extinction map obtained with [\fe] lines (see Figure\,\ref{fig:av_fe}), this latter cannot be used to
de-redden the H$_2$ emission. Hence, as a first step of the H$_2$ line analysis, we have derived 
a new extinction map.
Out of a number of ratios of lines coming from the same upper level, we considered only the three ratios 
(1-0S(1)/1-0Q(3), 2-0S(1)/2-1S(1), 2-0Q(3)/2-1S(1)) of lines detected at
S/N per pixel larger than 5 over the whole emission region. 
Since the one at the largest S/N (1-0S(1)/1-0Q(3)), suffers from the poor atmospheric 
transmission at 2.42$\mu$m of the 1-0Q(3), we have used 
the other two ratios as calibrators and the 1-0S(1)/1-0Q(3) ratio to probe the differential extinction 
along the shocked region. The Rieke \& Lebofsky (1985) extinction law was adopted. 
The final map is shown in Figure~\ref{fig:av_h2}: the covered zone 
somewhat complements the A$_V$ map constructed from [\fe] lines, with a partial spatial overlap in the areas
corresponding to knots B3 and B1 of D99. The A$_V$ values typically range from 1 to 4 mag,
in substantial agreement with those inferred from [\fe] emission: 
therefore this result demonstrates that the reddening {\it within} the shock does not change significantly
and that the A$_V$ variations mainly arise from the foreground material inside the cloud.
Our result contrasts with the general trend observed both in other Herbig-Haro objects (e.g.
Nisini et al., 2002, Giannini et al., 2004) and in HH99B itself (MC04), where the extinction 
computed from [\fe] lines is systematically higher than that probed with H$_2$ lines. This may stem because
in these cases the extinction was computed from the 1.257$\mu$m/1.644$\mu$m ratio, for which the {\it A} 
coefficients of NS have been adopted (see Sect.3.1.1).

\subsubsection{Temperature map}
\begin{figure*}
\centering
\includegraphics[width=18cm]{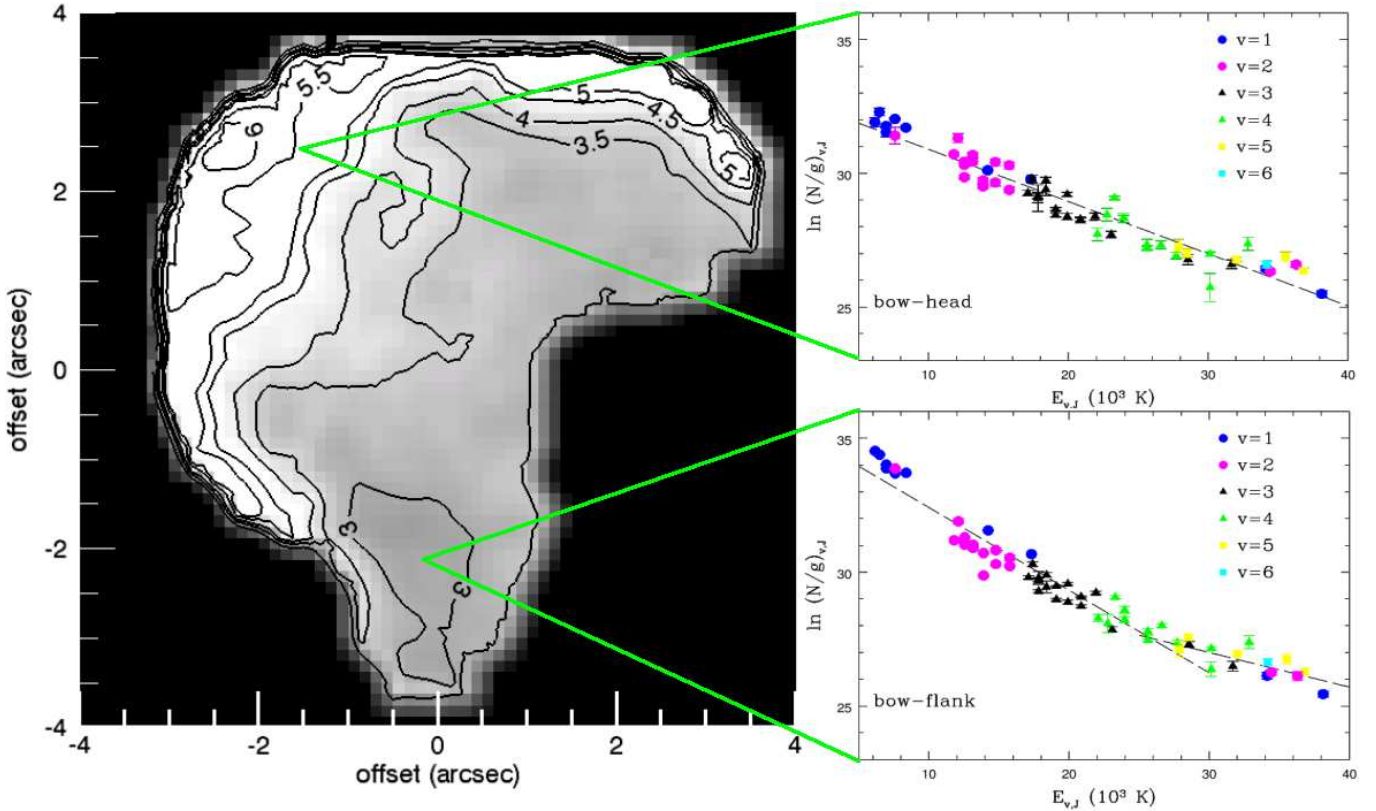}
\caption{\label{fig:temp_h2} Temperature map (in 10$^3$ K) as derived from 
H$_2$ line ratios. As an example, we show the rotational diagrams in two points
of the bow: while at the bow-head H$_2$ appears thermalised at
(T$\sim$ 5000 K), in the southern flank two temperature components co-exist, with an
average temperature (corresponding to the point shown in the map) of $\sim$ 2800 K.}
\end{figure*}

The temperature of the molecular gas can be obtained from a 
Boltzmann diagram (e.g. Gredel, 1994), plotting ln($N_{v,J}/g_J$) against $E_{v,J}/k$. Here
$N_{v,J}$ (cm$^{-2}$) is the column density of the level $(v,J)$,$E_{v,J}$ (K) its
excitation energy and  $g_J=(2J+1)(2I+1)$ the statistical weigth (here we assume that I=0,1
for para- and ortho-H$_2$). If
the gas is thermalised at a single temperature, the data points align onto
a straight line, whose slope gives the gas
temperature. In this diagram, points that refer to transitions from the same
upper level do overlay each other, once corrected for extinction effects.
We applied this method to all the pixels of the H$_2$ images, and the resulting
map is shown in Figure~\ref{fig:temp_h2}, where contours of temperature are given in units
of 10$^3$ K. 
Three results can be taken from this map: {\it (i)}
a temperature gradient from $\simeq$ 2\,000 K up to 6\.000 K occurs from the receding parts
of the shock towards the head. We underline that this gradient can be traced because
of the very large number of H$_2$ lines detected, which cover the Boltzmann diagram up 
to excitation energies of $\sim$ 38~000 K, therefore sensitively enlarging the dynamical 
range of temperatures typically probed with H$_2$ near-infrared lines; {\it (ii)} two different behaviours
in the Boltzmann diagram occur between the bow-head and the flanks (see the 
insets in Figure~\ref{fig:temp_h2}): H$_2$ appears fully thermalised (at T$\sim$
5000 K) at the bow-head (hence here the contours give directly the gas temperature), 
while a curvature exists among the points in the southern flank diagram, 
so that at least two temperature components can be traced. In these parts of the bow the 
contours indicate the {\it average} of these temperatures. 
The outlined behaviour can be generalised to the whole bow structure and likely reflects, 
from one side, that fluorescence can be discarded as a possible excitation mechanism, since it
implies strong departures from thermalisation (Black \& van Dishoeck, 1997), and, from the other
side, that  different shock mechanisms are at work. In
fact, the response of the level populations to the shock parameters can be seen in the 
Boltzmann diagram. Roughly speaking, an enhancement of the shock velocity 
increases the rate of collisions that vibrationally excite the H$_2$ molecules and favours 
thermodynamical equilibrium: thus the thermalisation observed along the head would
testifies the presence of a fast shock. If this were the case, a C-type shock should be favoured,
otherwise H$_2$ could not survive (at high temperatures) against dissociation 
(e.g. Le Bourlot et al., 2002, Flower et al., 2003). The alternative possibility that
the observed H$_2$ emission arises from re-forming H$_2$ onto dust grains can be reasonably 
discarded. In fact, the timescale of this process is of the order of 10$^{17}$/n s$^{-1}$ 
(Hollenbach \& McKee, 1979), being $n$ the density of the gas after the compression due to the 
shock passage. For reasonable values of this parameter (10$^4$-10$^7$ cm$^{-3}$), and assuming a 
velocity of the downstream gas of the order of 10 km s$^{-1}$, the distance covered by the gas at
the bow-head should be between 10$^{16}$ and 10$^{19}$ cm, definitively longer than the linear
dimension of knot B0, i.e. $\sim$ 5 10$^{15}$ cm; {\it (iii)} from the same Boltmann diagrams
a marked difference in the column density of the warm H$_2$ (from the intercept of the
straight line) emerges. For example, from the insets of Figure\,\ref{fig:temp_h2}
we derive N(H$_2$) = 2 10$^{17}$ cm$^{-2}$  and 3 10$^{16}$ cm$^{-2}$ at the flank and at the bow-head,
respectively. If the same shock lengths are assumed for these regions, it derives that about 85\% of 
the molecular gas must get destroyed in the most excited part of the shock (although H$_2$ emission
is still observed).

\subsection{H analysis}
At the bow-head we observe hydrogen recombination lines of the Brackett series along with
the Pa$\beta$ line. The S/N ratio of these lines, but Br$\gamma$
and Pa$\beta$, is so low, however, that detailed modelling is prohibitive.
Thus, we have used just the Pa$\beta$/Br$\gamma$ ratio to derive an
independent estimate of the extinction. Assuming case 
B recombination (Storey \& Hummer, 1995), we obtain A$_V$ = 1.8$\pm$1.9~mag.\\
More interesting parameters, i.e. the hydrogen fractional ionisation, x$_e$, and the hydrogen
post-shock density, n$_H$ = n$_e$/x$_e$, are obtainable from the observed intensity ratio 
[\fe]1.257/Pa$\beta$. As described in detail in Nisini et al.(2002), under the assumption
that iron is singly ionised, such a ratio can be expressed as:

\begin{equation}
 x_e= \delta_{Fe} (\Fe/\Hy)_{\odot} \left( \frac{[\fe]1.257}{Pa\beta} \right)^{-1}  \frac{\epsilon_{[\fe]1.257}}{\epsilon_{Pa\beta}}
\end{equation}

with $\delta_{Fe}$ the gas-phase iron fraction with respect to the solar \Fe\, abundance, 
(\Fe/\Hy)$_{\odot}$ and $\epsilon_{[\fe]1.257}$ and $\epsilon_{Pa\beta}$ (in erg cm$^{3}$ s$^{-1}$) 
the emissivities of the two lines, taking $\epsilon_{Pa\beta}$ from Storey \& Hummer (1995).
The above quantity was computed for the physical conditions derived at the
bow-head, i.e. T=16~000 K, n$_e$=6 10$^3$ cm$^{-3}$, $\delta_{Fe}$ $\sim$ 0.7 and 
having taken (\Fe/\Hy)$_{\odot}$ = 2.8\,10$^{-5}$ (Asplund, Grevesse \& Sauval 2005). This
leads to x$_e$ $\sim$ 0.4-0.5 and n$_H$ between 0.8 and 1.4 10$^4$ cm$^{-3}$. 
In the receding parts of the shocks, the fractional ionisation 
cannot be directly computed, since we have no estimates for the electron
temperature. However, under the reasonable assumption that T$_e$$\la$ 10 000 K
a sharp decrease of x$_e$ is expected: for example, for the regions where  
$\delta_{Fe}$=0.3-0.4, we find x$_e$ $\sim$ 0.2-0.3.
These estimates are in agreement with those inferred along
other Herbig-Haro objects through optical line diagnostics (e.g. Bacciotti \& Eisl\"{o}ffel,
1996, Hartigan \& Morse, 2007) and with those inferred in 
a number of jets by combining optical and infrared observations (Nisini et al., 2005, 
Podio et al., 2006). 

\section{Kinematical properties}
\subsection{H$_2$}

\begin{figure*}
\centering
\includegraphics[width=15cm]{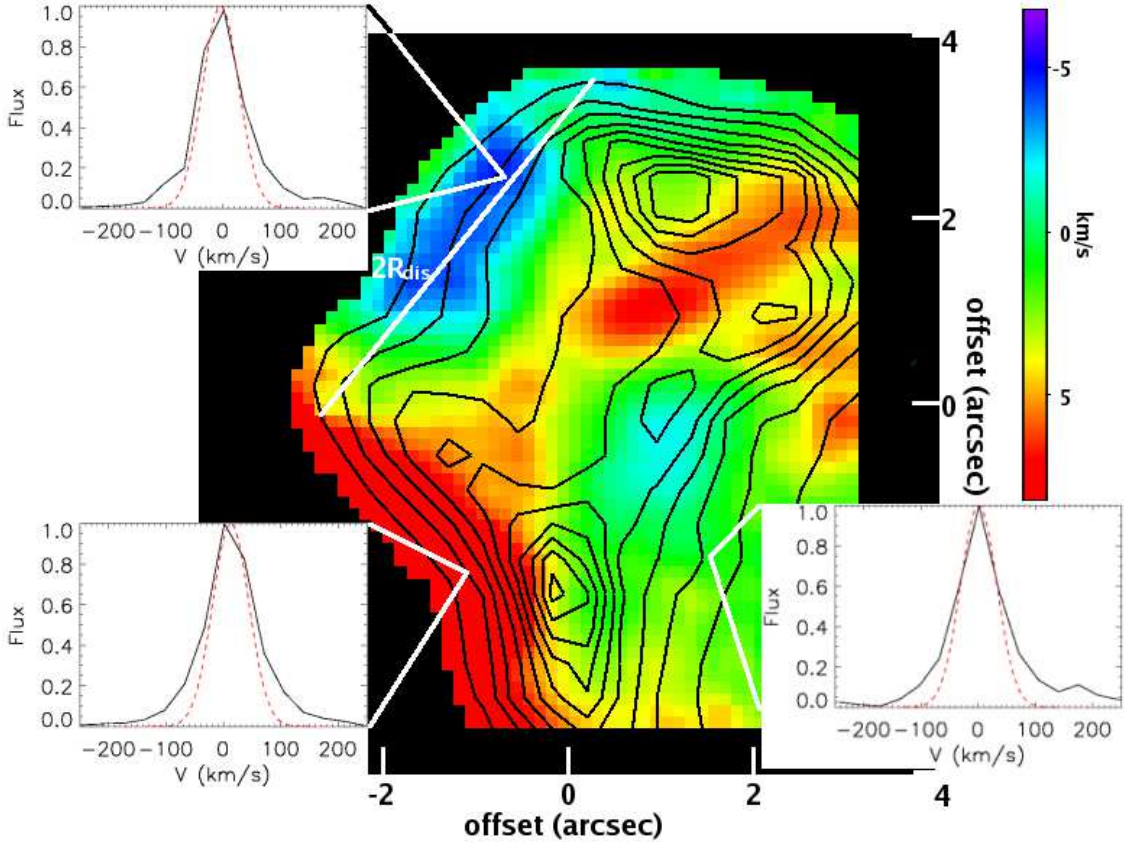}
\caption{\label{fig:vel_h2} Local standard of rest (LSR) 
velocity map of the 2.122$\mu$m line peak, with superimposed line intensity 
contours. We indicate with a white line the diameter 2R$_{dis}$ of the last cap beyond which the bulk of H$_2$
is dissociated (see Sect.4.2). Insets show the line profile observed at the bow head (top left), in the southern 
flank (bottom left) and at the bow centre (bottom right). Blue and red asymmetries are visible 
at the bow head and along the flanks, while the line is symmetric toward the centre.
The instrumental profile, measured on OH atmospheric lines, is shown for comparison (red dashed line).}
\end{figure*}

In this section we intend to characterise the kinematical parameters of the shock(s)
in HH99B. This topic was already discussed by D99, and we compare here our results 
with those found in that work. The higher spectral resolution (R$\sim$ 15
kms$^{-1}$) of D99, obtained with echelle spectroscopy, has revealed
that the peak velocity of the H$_2$ 2.122$\mu$m line moves progressively from slightly 
blue-shifted values near the shock front towards red-shifted values in the flanks, and has
been interpreted in the framework of a receding bow-shock oriented with respect 
to the line of sight of about 45$^{\circ}$.\\

Nominally, the spectral resolution of our K band observations would not permit us to reveal 
variations of the order of those measured by D99. This limitation, however, is partially 
compensated by the very high S/N ratio at which we detect the 2.122$\mu$m line. 
A trend on both the peak velocity and on the profile shape can be followed
along the bow structure, although we cannot give precise numerical estimates on the line 
parameters (v$_{peak}$, FWHM). Our results are presented in Figure\,\ref{fig:vel_h2}, where the contours 
of the 2.122$\mu$m line intensity (de-reddened) have been superposed on the v$_{peak}$ map.

Overall, our results confirm those of D99: the line profile 
presents a blue-shifted component towards the shock front at the bow head (B0). The opposite 
occurs along the two flanks and especially along the edge of the B1 flank 
(not covered by the echelle spectra in D99), where the line peak is shifted of 
$\sim$ +15 km s$^{-1}$ with respect to the line profile at the head. Analogous to the spectra of D99,
the 2.122$\mu$m profile does not show double peaked components, as generally expected for a 
parabolic bow structure (Schultz, Burton \& Brand, 2005) though it does become wider 
near the centre of the bow, where the opposite sides are seen 
in projection. Here the observed FHWM$_{obs}$ is 85-105 km s$^{-1}$, that, deconvolved with the
instrumental profile, measured on atmospheric OH lines, roughly gives an intrinsic line 
width of $\sim$ 20-40 km s$^{-1}$.
The agreement with previous observations is also mantained along the
bow flanks, where the profile width becomes narrower decreasing toward 
the spectral resolution limit on the intrisic width of $\sim$ 20 km s$^{-1}$. As for the line peak, 
a sudden increase of FHWM$_{obs}$ is registered at the edge of the southern flank, 
where we measure up to $\sim$ 115 km
s$^{-1}$, i.e. an intrinsic width of $\sim$ 70 km s$^{-1}$. This is close to the maximum shock velocity 
($\sim$ 80 km s$^{-1}$) at which H$_2$ can survive against dissociation, predicted by the 
C-shock model by Le Bourlot et al. (2002). This last topic will be further discussed in
the next section.

\subsection{[\fe]}
\begin{figure*}
\centering
\includegraphics[width=14cm]{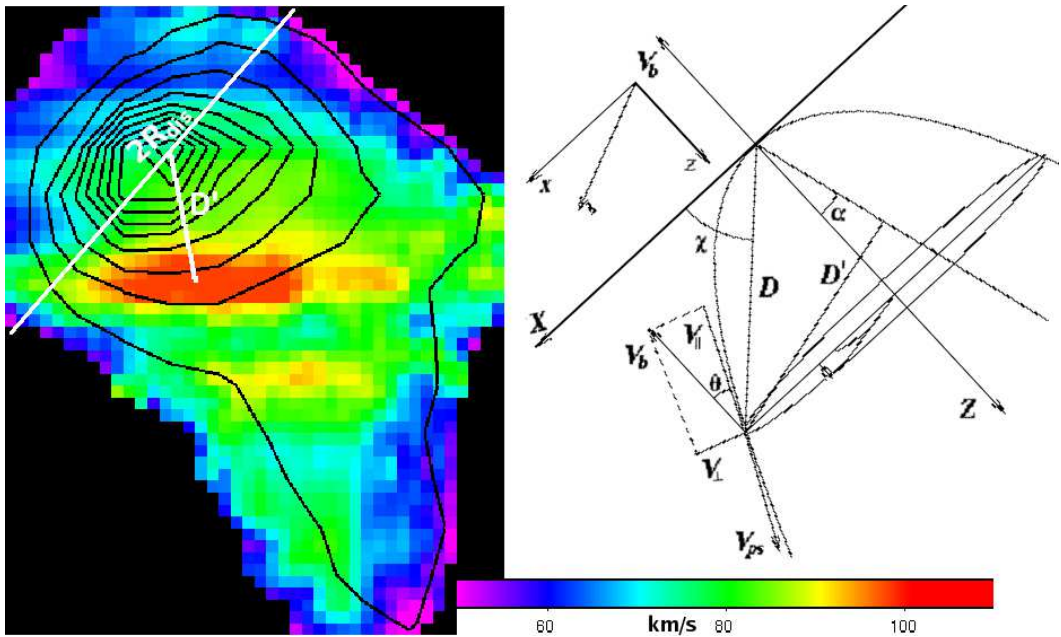}
\caption{\label{fig:vel_fe}{\it Left panel}: local standard of rest (LSR) 
velocity map of the 1.644$\mu$m line peak, with superimposed line intensity 
contours. The line is red-shifted along the bow-structure, with a peak at about +70 km s$^{-1}$.
The projection $D^{\prime}$ along the line of sight of the distance between the line emission peak and the
maximum radial velocity is shown, together with the diameter 2R$_{dis}$ (see caption of
Figure\,\ref{fig:vel_h2}). {\it Right panel}: schematic view of the parabolic geometry assumed for the
bow-shock. The apex coincides with the origin of the coordinate system. R=$\sqrt{x^2 + y^2}$, is the 
radius at the distance $z$ along the bow axis, $\phi$ is the azimuthal angle around
the z-axis, $\theta$ is the angle between the bow direction and the tangent to the parabolic surface 
(at each  $\theta$ the shock velocity at the apex, v$_b$, is de-composed in v$_{\parallel}$ and
v$_{\perp}$) and $\alpha$ is the angle, lying on the x-z plane, between the z-axis and the line of sight.
D is the distance between the line emission peak and the maximum radial velocity and $D^{\prime}$ its 
projection onto the sky plane.}
\end{figure*}
\begin{figure*}
\centering
\includegraphics[width=12cm]{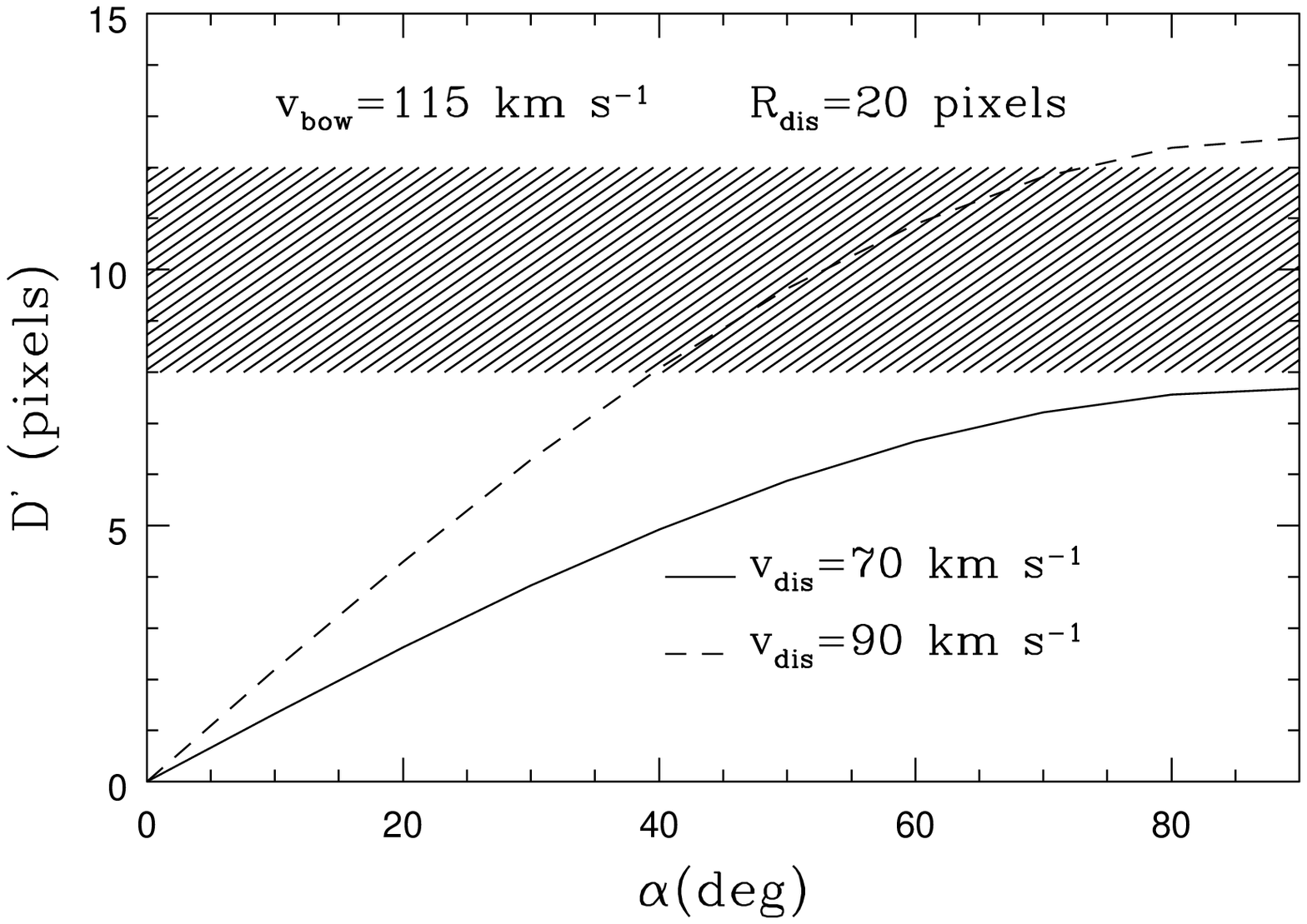}
\caption{\label{fig:alpha_plot} Projection over the sky plane of the distance between the
[\fe]1.644$\mu$m emission peak and the maximum radial velocity, plotted against the bow inclination angle
$\alpha$. The hatched area defines the uncertainty on $D^{\prime}$, estimated as the maximum and 
minimum distance between the 1.644$\mu$m line peak and the points where the maximum shift in velocity is 
registered (those coloured in red in Figure\,\ref{fig:vel_fe}, left panel).
The curves refer to different assumptions for the H$_2$ breakdown
speed, v$_{dis}$, having measured the shock velocity at the bow apex, v$_{bow}$ and the 
radius R$_{dis}$ (see Appendix A). The observations are matched for v$_{dis}$ between 70 and 90 km s$^{-1}$.}
\end{figure*}

We have performed the kinematical analysis of the ionic gas component on the two brightest
[\fe] lines at 1.257 and 1.644$\mu$m. Both appear resolved in velocity and give similar results
for the line profile shape (which is Gaussian across the whole bow), the intrinsic line width, 
the peak position and the FWZI. This last result gives a direct measure of the shock
speed at the bow apex (Hartigan, Raymond \& Hartmann, 1987): we obtain v$_{bow}$ = 115$\pm$5 km s$^{-1}$,
which agrees well with the prediction by D99 (80-120 km s$^{-1}$) derived from the overall shape
of the bow and the characteristics of the H$_2$\,2.122$\mu$m line profile.\\   
In Figure\,\ref{fig:vel_fe} we show the radial velocity map of the 1.644$\mu$m line: this appears 
red-shifted over the whole bow structure, the maximum shift occurring towards the image centre.
We intepret this behaviour as a geometrical effect due to the inclination of the bow with respect to
the line of sight. Indeed, if the bow is observed at a certain angle $\alpha$ $\ne$ 0$^{\circ}$,
180$^{\circ}$, the peak of the radial velocity component is seen offset from the bow apex. 
Following the procedure described in 
Appendix A, we are able to express the projected
distance (D$^{\prime}$) between the emission and the radial velocity peaks as a function of $\alpha$
and the H$_2$ breakdown velocity (v$_{dis}$), once the relationship between this latter
and the parameter $b$ regulating the bow (parabolic) shape is defined. 
The result is depicted in Figure \ref{fig:alpha_plot}: interestingly, we found $\alpha$ values
close to that inferred by D99 for a range of v$_{dis}$    
between 70 and 90 km s$^{-1}$. These values can be considered as an indirect measurement of a 
parameter whose theoretical predictions have been widely discussed over the last decades: a first value
of 24 km s$^{-1}$ was inferred by Kwan (1977), subsequently Draine, Roberge \& Dalgarno (1983) found 
v$_{dis}$=50 km s$^{-1}$, a result confirmed by Smith \& Brand (1990) and Smith (1991). More recently,
Le Bourlot et al. (2002) have shown that v$_{dis}$ can increase up to 80 km s$^{-1}$ for values of the
pre-shock density around 10$^3$ cm$^{-3}$. Our measurement is therefore in agreement with this latter
prediction. Moreover, v$_{dis}$ $\sim$ 80 km s$^{-1}$ is consistent with the FWHM of the 2.122$\mu$m line 
profile measured in the southern flank (Sect. 4.1), and reinforces the hypothesis that a fast, continuous 
shock is responsible for the excitation of the molecular gas at the apex of the bow (Sect.3.2.2).

\begin{table*}
\caption{\label{tab:tab4} Physical parameters estimated in HH99B.}\small
\begin{center}
\begin{tabular}{lccc}
\hline\\[-5pt]
Parameter                       &  This work &  MC04   & D99 \\
\hline
A$_V$(\fe) (mag)                  & 1-4        &       4-10        &       -       \\
A$_V$(H$_2$) (mag)               & 2-4        &        $<$4        &      $<$4.6   \\
A$_V$(H) (mag)                  & $\sim$2     &         -         &        -      \\
n$_e$  (cm$^{-3}$)               & 3-6 10$^3$  &         -         &      -        \\
T$_e$ (K)                       & $\le$ 17~000&      -            &       -       \\
T$_{H_2}$ (K)                   & 2~500 - 6~000&    2~000 - 4~000 &      -        \\
N$_{H_2}$ (cm$^{-2}$)            & 3 10$^{16}$ - 2 10$^{17}$        &       -        \\
$\delta$$_{\fe}$                 & 0.2-0.7    &  0.25             &       -        \\
x$_e$ 		                & 0.2-0.5    &       -           &      -         \\
n$_H$  (cm$^{-3}$)               & $\sim$10$^4$ (post-shock) &  10$^4$ (pre-shock)  &       -         \\
v$_{bow}$ (km s$^{-1}$)            & 110-120 ([\fe])  &    50 ([\fe])      &     80-120 (H$_2$)    \\
inclination angle ($^{\circ}$)   &  40-60    &   -              &     45          \\
\hline\\[-5pt]
\end{tabular}
\end{center} 
\end{table*}

\section{Concluding remarks}

We have presented bi-dimensional, deep near-infrared spectral images of the bow shock
HH99B. These have allowed us, for the first time, to accurately derive the 
physical parameters of both the molecular and ionic gas components (summarised in
Table\,\ref{tab:tab4}, where our results are compared with those derived in previous works), and, 
at the same time, to characterise the geometry and the 
kinematical properties of the flow. 
The main results of this study are the following:
\begin{itemize}
\item[-] More than 170 emission lines have been detected, mainly ro-vibrational H$_2$
and [\fe] lines, many of them never observed before in an Herbig-Haro object.
In addition, transitions of hydrogen and helium recombination and fine
structure lines of [\pii], [\tiii], and possibly [\coii] have been observed. 
\item[-] A clear bow-shape morphology emerges from the line intensity maps. As shown in Figures\,
\ref{fig:spettro_j}-\ref{fig:spettro_k}, [\fe] and other
ionic emission peaks definitely at the bow-head (B0), being strong also in the knot B2
immediately behind. In contrast, H$_2$ emission delineates the bow flanks, peaking in the
knots B1 and B3. Notably, the H$_2$ lines with the highest excitation
energy (E$_{up}$ $\ga$ 30\,000 K) show a different morphology, being strong towards the bow-head. 
This implies that H$_2$ still survives in this zone, in spite of the significant temperature 
enhancement there.
\item[-] Extinction maps have been derived from the analysis of both [\fe] and H$_2$ lines.
These give similar results, with A$_V$ ranging between 1 and 4 mag.
\item[-] A detailed electron density map has been obtained in the framework of NLTE approximation
for [\fe] line emission. This remains almost constant in the [\fe] emission zone, peaking towards
the bow-head. From the same emission, we are able to probe a variation of the electron temperature,
which falls from $\sim$ 16\,000 K at the apex to less than 10\,000 K in the receding parts of the bow.
\item[-] An iron depletion degree not higher than 30\% has been inferred at the bow apex, which testifies
in favour of a J-type shock as the main excitation mechanism in this part of the bow. In this same zone,
we infer a fractional ionisation of $\sim$ 0.6 and a post-shock density of $\sim$ 10$^4$ cm$^{-3}$.
\item[-] Analysis of H$_2$ line emission has allowed us to probe the molecular temperature variation.
We find that at the bow apex thermalisation has been reached at T$\sim$ 6\,000 K, likely 
due to the presence of a fast, non-dissociative shock.
On the contrary, along the flanks different temperature components are simultaneously present. 
A decrease of about a factor of ten is registered in the H$_2$ column density going from knots B1/B3 toward knot B0.
Both these circumstances are not accounted for by models which have attempted to interpret
the H$_2$ emission on HH99B on the basis of fewer lines. Therefore, the conclusions of such models should 
be tested in the light of this new piece of information.
\item[-] From the brigthest [\fe] and H$_2$ lines we have been able to probe the kinematical properties
(e.g. shock velocity) of the shocked gas. In particular, we confirm the result by D99 according to which HH99 is a red-shifted,
receding bow. 
\item[-] The radial velocity map of [\fe] emission has been interpreted in the framework of the bow
geometry. From this map we have consistently inferred the bow inclination angle and defined a range of
70-90 km s$^{-1}$ for the H$_2$ breakdown velocity. These values are in agreement with the prediction
of Le Bourlot et al.(2002) of v$_{dis}$ up to 80 km s$^{-1}$. We propose our method as a valuable tool to 
derive the jet inclination angle (if larger than 10-20$^{\circ}$) in cases where proper motion
is unknown.
\item[-] The kinematical parameters of the [\fe] emission estimated in this work do not confirm
the model predictions by MC04. In particular, the argument that the [\fe] lines
originate in a pure J-type shock with $v_{shock}$ $\sim$ 50 km s$^{-1}$ contrasts 
with our measure of $v_{shock}$ $\sim$ 110-120 km s$^{-1}$. Thus, the modelling of 
the [\fe] emission may have to be revised.
\end{itemize}

\section{Acknowledgments}
We are grateful to P. Hartigan for his suggestions and comments about
the determination of the Einstein coefficients for iron.
We also acknowledge E. Oliva and M.A. Bautista for having providing us the
electronic collisional rates for phosphorus and titanium. 
This work was partially supported by the European
Community's Marie Curie Research and Training Network JETSET (Jet
Simulations, Experiments and Theory) under contract MRTN-CT-2004-005592.

\begin{appendix}
\section{Derivation of inclination angle and H$_2$ breakdown velocity}
With reference to Figure\,\ref{fig:vel_fe}, right panel, we assume that the bow has a parabolic geometry
($z=R^2/2b$), where the apex coincides with the origin of the coordinate system. R=$\sqrt{x^2 + y^2}$, is 
the radius at the distance $z$ along the bow axis, $\phi$ is the azimuthal angle around
the z-axis, $\theta$ is the angle between the bow direction and the tangent to the parabolic surface 
and $\alpha$ is the angle, lying on the x-z plane, between the z-axis and the line of sight.
This latter can be inferred by measuring in the radial velocity map (Figure\,\ref{fig:vel_fe}, left panel) 
the projection $D^{\prime}$ over the sky plane of the distance between the line emission peak and the 
maximum radial velocity (D$^{\prime}$ =10$\pm$2 pixels). We have:

\begin{eqnarray}
D^{\prime} = D \cos (\chi - \alpha)= \sqrt{z_{max}^2 + 2z_{max}b} \times \cos (\chi - \alpha)
\label{a1} 
\end{eqnarray}

where z$_{max}$  is the distance along the z-axis at which the radial velocity reaches its maximum.
We firstly estimated the $b$ parameter. Since the molecular hydrogen
does not emit over the entire bow surface, there is a leading edge which divides the bow into two
different zones: a dissociation cap, beyond which the emission comes from atomic/ionic elements and a molecular
hydrogen emitting flank. Therefore, it is possible to define as z$_{dis}$ the position along the z-axis 
where the leading edge plane is located: certainly it depends on the shock velocity at the bow apex,
v$_{bow}$, on the H$_2$ breakdown velocity, v$_{dis}$ and on the shape of the bow surface. Setting 
v$_{dis}$ = v$_{bow}$ sin$\theta_{dis}$ and, since for a generic angle $\theta$, tan$\theta$=dR/dz:

\begin{eqnarray}
v_{dis}= \left. v_{bow} \frac{dR/dz}{\sqrt{1+\left( \frac{dR}{dz}\right)^2}} \right|_{z=z_{dis}}
\label{a2}
\end{eqnarray}

Substituting the parabolic equation in this formula, we get:

\begin{eqnarray}
z_{dis} = \frac{b}{2}\left[ \left( \frac{v_{bow}}{v_{dis}} \right)^2 - 1 \right]
\label{a3}
\end{eqnarray}

and 

\begin{eqnarray}
b = \frac{R_{dis}^2}{2z_{dis}}  = \frac{R_{dis}}{\sqrt{\left( \frac{v_{bow}}{v_{dis}}\right)^2 - 1}}
\label{a4}
\end{eqnarray}

In eq.\ref{a4}, R$_{dis}$ is the radius of the cap beyond which H$_2$ dissociates, that we have 
measured both from the 2.122$\mu$m (Figure\,\ref{fig:vel_h2}) and  1.644$\mu$m emission contours
(Figure\,\ref{fig:vel_fe}). Both of them give R$_{dis}$ $\sim$ 20 pixels. We also have estimated
v$_{bow}$ = 115$\pm$5 km s$^{-1}$ from the FWZI of the 1.644$\mu$m (Hartigan, Raymond \& Hartmann, 
1987). Substituting these values in eq.\ref{a4}, we $b$ as a function of v$_{dis}$ only, 
that we have taken as a free parameter.

The above expression of $b$ is then substituted in the parabolic equation to derive z$_{max}$:

\begin{eqnarray}
z_{max} = \frac{b}{2}\left[ \left( \frac{v_{bow}}{v_{max}} \right)^2 - 1 \right] = \frac{b}{2}\left[\frac{1}{\sin\theta_{max}^2 -1} \right]
\label{a5}
\end{eqnarray}

Following Hartigan, Raymond \& Hartmann (1987) and considering that the bow is seen from the back
(D99), we get $\theta_{max}$ = $\pi$/2 - $\alpha$/2. Thus:

\begin{eqnarray}
z_{max} = \frac{b}{2}\left[ \left( \frac{1}{\cos {(\alpha/2)}} \right)^2 - 1 \right]
\label{a6}
\end{eqnarray}

In conclusion, from equations \ref{a4} and \ref{a6} and by measuring $\chi$ from simple geometrical 
considerations, we are able to express in eq.\ref{a1} D$^{\prime}$ as a function only of 
$\alpha$ and v$_{dis}$, that have been routinely varied to match 
our measurement of D$^{\prime}$, as shown in Figure\,\ref{fig:alpha_plot}.  
\end{appendix}

\begin{thebibliography}{}
\bibitem{} Asplund, M., Grevesse, N., \& Sauval, A.J. 2005, Cosmic Abundances as
Records of Stellar Evolution and Nucleosynthesis, 336, 25    
\bibitem{} Bacciotti, F., \&  Eisl\"{o}ffel, J. 1999, A\&A, 342, 717
\bibitem{} Black, J.H., \& van Dishoeck, E.F. 1997, ApJ, 322, 412 
\bibitem{} Beck-Winchatz, B., Bo\.hm, K-H, \& Noriega-Crespo, A. 1996, AJ, 111, 346
\bibitem{} Bonnet, H., Abuter, R., Baker, A., et al. 2004, The ESO Messenger 117, 17
\bibitem{} Caratti o Garatti, A., Eisl\"{o}ffel, J., Froebrich, D., et al. 2007, A\&A, submitted 
\bibitem{} Cardelli, J. A., Clayton, G. C.\& Mathis, J. S. 1988, ApJ, 329,33
\bibitem{} Calzoletti, L., Giannini, T., Nisini B., et al., 2008, in preparation
\bibitem{} Draine, B.T. 1980, ApJ, 241, 1021 
\bibitem{} Draine, B.T., Roberge, W.G., \& Dalgarno, A. 1983, ApJ, 264, 485
\bibitem{} Davis, C.J., Smith, M.D., Eisl\"{o}ffel, J., Davies, J.K. 1999, MNRAS, 308, 539 (D99)
\bibitem{} Eisenhauer, F., Abuter, R., Bickert, K. et al., 2003, SPIE 4841, 1548
\bibitem{} Eisl\"{o}ffel, J., Smith, M.D., \& Davis, C.J. 2000, A\&A, 359, 1147 
\bibitem{} Flower, D.R., Le Bourlot, J., Pineau des Fore\^ts, G. Cabrit, S. 2003, MNRAS, 341, 70
\bibitem{} Garcia-Lopez, R., Nisini, B., Giannini, T., et al. 2008, A\&A, submitted
\bibitem{} Giannini, T., M$^c$Coey, C., Caratti o Garatti, A., Lorenzetti, D., Flower, D. 2004, A\&A, 419,
999 
\bibitem{} Giannini, T., M$^c$Coey, C., Nisini, B., et al. 2007, A\&A, 459, 821
\bibitem{} Gredel, G. 1994, A\&A, 292, 580
\bibitem{} Gredel, G., 2006 A\&A, 457, 157
\bibitem{} Jones, A.P. 2000, JGR, 105, 10257
\bibitem{} Hartigan, P. 1989, ApJ, 339, 987
\bibitem{} Hartigan, P., \& Graham, J.A. 1987, AJ, 93, 913
\bibitem{} Hartigan, P., \& Morse, J. 2007, ApJ, 660, 426
\bibitem{} Hartigan, P., Raymond J., \& Hartmann L. 1987, ApJ, 316, 323
\bibitem{} Hollenbach, D., \& McKee, C. 1979, ApJ, 41, 555
\bibitem{} Hollenbach, D., \& McKee, C. 1989, ApJ, 342, 306
\bibitem{} Kwan, J. 1977, ApJ, 216, 713
\bibitem{} Lesaffre, P., Chie\^ze, J.-P., Cabrit, S., Pineau des Fore\^ts, G. 2004, A\&A, 427,147(a)
\bibitem{} Lesaffre, P., Chie\^ze, J.-P., Cabrit, S., Pineau des Fore\^ts, G. 2004, A\&A, 427,157(b)
\bibitem{} Le Bourlot, J. Pineau des Fore\^ts, G. Flower, D.R., Cabrit, S. 2002, A\&A, 390, 369
\bibitem{} Lidman, C., \& Cuby, J.G. 2000, ATLAS of OH lines
\bibitem{} Marraco, H.G., \& Rydgren A.E. 1981, AJ, 86, 62 
\bibitem{} Mouri, H., \& Taniguchi, Y. 2000, ApJ, 534, L63
\bibitem{} May, P.W., Pineau des Fore\^ts,G., Flower, D.R., et al. 2000, MNRAS, 318, 809
\bibitem{} M$^c$Coey, C., Giannini, T., Flower, D.R., Caratti o Garatti, A. 2004, MNRAS, 353, 813 (MC04)
\bibitem{} Modigliani, A., Ballester, P. \& Peron, M. 2007, {\it SINFONI Pipeline User Manual},
http://www.eso.org/projects/dfs/dfs-shared/web/vlt/vlt-instrument-pipelines.html
\bibitem{} Nisini, B., Bacciotti, F., Giannini, T., et al. 2005 A\&A, 441, 159
\bibitem{} Nisini, B., Caratti o Garatti, A., Giannini, T., Lorenzetti, D. 2002, A\&A, 393, 637
\bibitem{} Nussbaumer, H., \& Storey, P.J. 1988, A\&A, 193, 327
\bibitem{} O'Connell,B., Smith, M.D., \& Davis, C.J. 2004, A\&A, 419, 475
\bibitem{} Oliva, E., Marconi, A., Maiolino, R., et al. 2001 A\&A, 369, 5
\bibitem{} Pradhan, A.K., \& Zhang, H.L. 1993, ApJ, 409, L77
\bibitem{} Podio, L., Bacciotti, F., Nisini, B., et al. 2006, A\&A, 456, 189
\bibitem{} Reipurth, B., \& Bally, J. 2001, ARAA, 39, 40
\bibitem{} Rieke, G.H., \& Lebofsky, M.J. 1985, ApJ, 288, 618
\bibitem{} Quinet, P., Le Dourneuf, M., \& Zeippen, C.J. 1996, A\&AS, 120, 361
\bibitem{} Schultz, A.S.B., Burton, M.G., \& Brand, P.W.J.L. 2005, MNRAS, 358, 1195
\bibitem{} Smith, M.D. 1991, MNRAS, 252, 378
\bibitem{} Smith, M.D., \& Brand, P.W.J.L. 1990, MNRAS, 248, 108
\bibitem{} Smith, M.D., Froebrich, D., \& Eisl\"{o}ffel, J. 2003, ApJ, 592, 245
\bibitem{} Smith, M.D., \& Mac Low, M.-M. 1997, A\&A, 326, 801
\bibitem{} Smith, M.D., \& Rosen, A. 2003, MNRAS, 339,133
\bibitem{} Smith, N., \& Hartigan, P. 2006, ApJ, 638, 1045
\bibitem{} Storey, P.J., \& Hummer, D.G. 1995, MNRAS, 272, 41
\bibitem{} Walmsley, C. M., Natta, A., Oliva, E., Testi, L. 2000, A\&A, 364, 301
\bibitem{} Wilking, B.A., McCaughrean, M. J., Burton, M.G. et al., 1997, AJ, 114, 2029
\bibitem{} Zhang, H. L., \& Pradhan, A. K. 1995, A\&A, 293, 953

\end{thebibliography}
\end{document}